%% file: MOM-arxiv.tex
\documentclass[zpreprint,zbstnp,final]{zeus_paper}
\usepackage{placeins}
\usepackage[titletoc,title]{appendix}
\usepackage{authblk}

\usepackage{hyperref}
\usepackage{calc}
\usepackage[utf8]{inputenc}
\usepackage{graphicx}
\usepackage{ifthen}
\usepackage{colortbl}
\usepackage{titlesec}
\usepackage{xspace}
\usepackage{etex}
\usepackage[export]{adjustbox}
\usepackage{setspace}
\newcommand{\BROADFIGWIDTH}{0.84\textwidth}
\newcommand{\NARROWFIGWIDTH}{0.75\textwidth}
\newcommand{\oneminusT}{\tau}
\graphicspath{{Figures/}{}}

\title{Determination of $\alpha_{S}$ beyond NNLO using event shape averages}
\author[a]{Adam Kardos\thanks{kardos.adam@science.unideb.hu}}
\author[b]{G\'abor Somogyi\thanks{somogyi.gabor@science.unideb.hu}}
\author[c]{Andrii Verbytskyi\thanks{andrii.verbytskyi@mpp.mpg.de}}
\affil[a]{University of Debrecen, 4010 Debrecen, PO Box 105, Hungary}
\affil[b]{ELKH-DE Particle Physics Research Group, University of Debrecen,\protect\\ 
4010 Debrecen, PO Box 105, Hungary}
\affil[c]{Max-Planck-Institut f\"{u}r Physik, 80805 Munich, Germany}

\prepnum{MPP-2020-167}

\newcommand{\epjconly}[1]{}
\newcommand{\draftonly}[1]{}
\newcommand{\arxivonly}[1]{#1}
\include{MOM-tab}

\include{MOM-val}
\include{MOM-def}

\begin{document}
\abstract{\input{MOM-abs}}
\makezeustitle 
\newpage
\clearpage
\pagenumbering{arabic}
\pagestyle{plain}
\include{MOM-txt}

%%%%%%%%%%%%%%%%%%%%%%%%%%%%%%%%%%%%%%%%%%%%%%%%%%%%%%%%%%%%%%%%%%%%%%%%
{\bibliographystyle{./MOM}{\raggedright\bibliography{MOM.bib}}}\vfill\eject
\clearpage
\end{document}

%% file: MOM-tab.tex
\newcommand{\MOMtabularinputoneminusT}{
ALEPH~\protect\cite{Buskulic:1996tt} &  $\langle \oneminusT^{1}\rangle$ & $1$,$[133]$&  $\langle \oneminusT ^{1}\rangle$ & $1$,$[133]$\\
ALEPH~\protect\cite{Buskulic:1992hq} &  $\langle \oneminusT^{1}\rangle$ & $1$,$[91]$&  $\langle \oneminusT ^{1}\rangle$ & $1$,$[91]$\\
ALEPH~\protect\cite{Heister:2003aj} &  $\langle \oneminusT^{1}\rangle$ & $9$,$[91,206]$&  $\langle \oneminusT ^{1}\rangle$ & $9$,$[91,206]$\\
AMY~\protect\cite{Li:1989sn} &  $\langle \oneminusT^{1}\rangle$ & $1$,$[55]$&  $\langle \oneminusT ^{1}\rangle$ & $1$,$[55]$\\
DELPHI~\protect\cite{Abreu:1999rc} &  $\langle \oneminusT^{1,2,3}\rangle$ & $15$,$[91,183]$&  $\langle \oneminusT ^{1}\rangle$ & $5$,$[91,183]$\\
DELPHI~\protect\cite{Reinhardt:2001zz} &  $\langle \oneminusT^{1}\rangle$ & $15$,$[45,202]$&  $\langle \oneminusT ^{1}\rangle$ & $11$,$[45,202]$\\
HRS~\protect\cite{Bender:1984fp} &  $\langle \oneminusT^{1}\rangle$ & $1$,$[29]$&  $\langle \oneminusT ^{1}\rangle$ & $1$,$[29]$\\
JADE~\protect\cite{Pahl:2007zz} &  $\langle \oneminusT^{1,2,3,4,5}\rangle$ & $30$,$[14,43]$&  $\langle \oneminusT ^{1}\rangle$ & $4$,$[34,43]$\\
L3~\protect\cite{Adeva:1992gv} &  $\langle \oneminusT^{1}\rangle$ & $1$,$[91]$&  $\langle \oneminusT ^{1}\rangle$ & $1$,$[91]$\\
L3~\protect\cite{Achard:2004sv} &  $\langle \oneminusT^{1,2}\rangle$ & $30$,$[41,206]$&  $\langle \oneminusT ^{1}\rangle$ & $15$,$[41,206]$\\
MARK~\protect\cite{Abrams:1989ez} &  $\langle \oneminusT^{1}\rangle$ & $1$,$[89]$&  $\langle \oneminusT ^{1}\rangle$ & $1$,$[89]$\\
MARK~\protect\cite{Petersen:1987bq} &  $\langle \oneminusT^{1}\rangle$ & $1$,$[29]$&  $\langle \oneminusT ^{1}\rangle$ & $1$,$[29]$\\
MARKII~\protect\cite{Abrams:1989ez} &  $\langle \oneminusT^{1}\rangle$ & $1$,$[89]$&  $\langle \oneminusT ^{1}\rangle$ & $1$,$[89]$\\
OPAL~\protect\cite{Pahl:2007zz} &  $\langle \oneminusT^{1,2,3,4,5}\rangle$ & $60$,$[91,206]$&  $\langle \oneminusT ^{1}\rangle$ & $12$,$[91,206]$\\
TASSO~\protect\cite{Braunschweig:1990yd} &  $\langle \oneminusT^{1}\rangle$ & $4$,$[14,44]$&  $\langle \oneminusT ^{1}\rangle$ & $2$,$[35,44]$\\
}
\newcommand{\MOMtabularinputC}{
ALEPH~\protect\cite{Buskulic:1992hq} &  $\langle C^{1}\rangle$ & $1$,$[91]$&  $\langle C ^{1}\rangle$ & $1$,$[91]$\\
DELPHI~\protect\cite{Reinhardt:2001zz} &  $\langle C^{1}\rangle$ & $15$,$[45,202]$&  $\langle C ^{1}\rangle$ & $11$,$[45,202]$\\
DELPHI~\protect\cite{Abreu:1999rc} &  $\langle C^{1,2,3}\rangle$ & $12$,$[133,183]$&  $\langle C ^{1}\rangle$ & $4$,$[133,183]$\\
JADE~\protect\cite{Pahl:2007zz} &  $\langle C^{1,2,3,4,5}\rangle$ & $30$,$[14,43]$&  $\langle C ^{1}\rangle$ & $4$,$[34,43]$\\
L3~\protect\cite{Adeva:1992gv} &  $\langle C^{1}\rangle$ & $1$,$[91]$&  $\langle C ^{1}\rangle$ & $1$,$[91]$\\
L3~\protect\cite{Achard:2004sv} &  $\langle C^{1,2}\rangle$ & $18$,$[130,206]$&  $\langle C ^{1}\rangle$ & $9$,$[130,206]$\\
OPAL~\protect\cite{Pahl:2007zz} &  $\langle C^{1,2,3,4,5}\rangle$ & $60$,$[91,206]$&  $\langle C ^{1}\rangle$ & $12$,$[91,206]$\\
}

%% file: MOM-val.tex
\newcommand{\TTMOMresultNNNLO}{0.14092\pm 0.00116 {\text( exp.)}\pm0.00111 {\text( hadr.)}\pm0.0090 {\text( scale)}}
\newcommand{\TTMOMresultNNNLOpII}{0.00111 {\text( hadr.)}\pm0.0090 {\text( scale)}}
\newcommand{\TTMOMresultNNNLOpI}{0.14092\pm 0.00116 {\text( exp.)}}
\newcommand{\TTMOMresultNNNLOquality}{79.2/63}
\newcommand{\TTDoneminusTpowone}{-7.51\times10^{4}\pm 1.14\times10^{3} {\text( exp.)}}
\newcommand{\TTMOMresultNNLO}{0.11459\pm 0.00022 {\text( exp.)}\pm0.00024 {\text( hadr.)}\pm0.0025 {\text( scale)}}
\newcommand{\TTMOMresultNNLOpII}{0.00024 {\text( hadr.)}\pm0.0025 {\text( scale)}}
\newcommand{\TTMOMresultNNLOpI}{0.11459\pm 0.00022 {\text( exp.)}}
\newcommand{\TTMOMresultNNLOquality}{324.3/64}
\newcommand{\TTBMOMresultNNNLO}{0.13011\pm 0.00171 {\text( exp.)}\pm 0.0137 {\text( scale)}}
\newcommand{\TTBMOMresultNNNLOpII}{0.0137 {\text( scale)}}
\newcommand{\TTBMOMresultNNNLOpI}{0.13011\pm 0.00171 {\text( exp.)}}
\newcommand{\TTBMOMresultNNNLOquality}{76.3/61}
\newcommand{\TTBMOMresultNNNLOalphazero}{0.91\pm 0.07 {\text( exp.)} \pm 0.31 {(scale)}}
\newcommand{\TTBMOMresultNNNLOMilan}{1.50\pm 0.17 {\text( exp.)}}
\newcommand{\TTBDoneminusTpowone}{-9.87\times10^{4}\pm 1.26\times10^{4} {\text( exp.)}}
\newcommand{\TTBMOMresultNNLO}{0.11940\pm 0.00125 {\text( exp.)}\pm 0.0019 {\text( scale)}}
\newcommand{\TTBMOMresultNNLOpII}{0.0019 {\text( scale)}}

\newcommand{\TTBMOMresultNNLOquality}{77.1/62}
\newcommand{\TTBMOMresultNNLOalphazero}{0.47\pm 0.03 {\text( exp.)} \pm 0.025 {(scale)}}
\newcommand{\TTBMOMresultNNLOMilan}{1.53\pm 0.30 {\text( exp.)}}
\newcommand{\TTbMOMresultNNNLO}{0.12911\pm 0.00177 {\text( exp.)}\pm 0.0123 {\text( scale)}}
\newcommand{\TTbMOMresultNNNLOpII}{0.0123 {\text( scale)}}
\newcommand{\TTbMOMresultNNNLOpI}{0.12911\pm 0.00177 {\text( exp.)}}
\newcommand{\TTbMOMresultNNNLOquality}{76.3/61}
\newcommand{\TTbMOMresultNNNLOalphazero}{0.89\pm 0.07 {\text( exp.)} \pm 0.29 {(scale)}}
\newcommand{\TTbMOMresultNNNLOMilan}{1.50\pm 0.17 {\text( exp.)}}
\newcommand{\TTbDoneminusTpowone}{-9.36\times10^{4}\pm 1.33\times10^{4} {\text( exp.)}}
\newcommand{\TTbMOMresultNNLO}{0.11927\pm 0.00125 {\text( exp.)}\pm 0.0018 {\text( scale)}}
\newcommand{\TTbMOMresultNNLOpII}{0.0018 {\text( scale)}}
\newcommand{\TTbMOMresultNNLOpI}{0.11927\pm 0.00125 {\text( exp.)}}
\newcommand{\TTbMOMresultNNLOquality}{77.0/62}
\newcommand{\TTbMOMresultNNLOalphazero}{0.48\pm 0.03 {\text( exp.)} \pm 0.02 {(scale)}}
\newcommand{\TTbMOMresultNNLOMilan}{1.53\pm 0.30 {\text( exp.)}}
\newcommand{\TTYMOMresultNNNLO}{0.12665\pm 0.00193 {\text( exp.)}\pm 0.0096 {\text( scale)}}
\newcommand{\TTYMOMresultNNNLOpII}{0.0096 {\text( scale)}}
\newcommand{\TTYMOMresultNNNLOpI}{0.12665\pm 0.00193 {\text( exp.)}}
\newcommand{\TTYMOMresultNNNLOquality}{76.3/61}
\newcommand{\TTYMOMresultNNNLOalphazero}{0.84\pm 0.07 {\text( exp.)} \pm 0.23 {(scale)}}
\newcommand{\TTYMOMresultNNNLOMilan}{1.50\pm 0.19 {\text( exp.)}}
\newcommand{\TTYDoneminusTpowone}{-7.97\times10^{4}\pm 1.52\times10^{4} {\text( exp.)}}
\newcommand{\TTYMOMresultNNLO}{0.11906\pm 0.00126 {\text( exp.)}\pm 0.0017 {\text( scale)}}
\newcommand{\TTYMOMresultNNLOpII}{0.0017 {\text( scale)}}
\newcommand{\TTYMOMresultNNLOpI}{0.11906\pm 0.00126 {\text( exp.)}}
\newcommand{\TTYMOMresultNNLOquality}{76.9/62}
\newcommand{\TTYMOMresultNNLOalphazero}{0.48\pm 0.03 {\text( exp.)} \pm 0.02 {(scale)}}
\newcommand{\TTYMOMresultNNLOMilan}{1.53\pm 0.29 {\text( exp.)}}
\newcommand{\CCMOMresultNNNLO}{0.14120\pm 0.00096 {\text( exp.)}\pm0.00097 {\text( hadr.)}\pm0.0100 {\text( scale)}}
\newcommand{\CCMOMresultNNNLOpII}{0.00097 {\text( hadr.)}\pm0.0100 {\text( scale)}}
\newcommand{\CCMOMresultNNNLOpI}{0.14120\pm 0.00096 {\text( exp.)}}
\newcommand{\CCMOMresultNNNLOquality}{40.8/40}
\newcommand{\CCDCpowone}{-3.10\times10^{5}\pm 3.21\times10^{3} {\text( exp.)}}
\newcommand{\CCMOMresultNNLO}{0.11298\pm 0.00020 {\text( exp.)}\pm0.00019 {\text( hadr.)}\pm0.0022 {\text( scale)}}
\newcommand{\CCMOMresultNNLOpII}{0.00019 {\text( hadr.)}\pm0.0022 {\text( scale)}}
\newcommand{\CCMOMresultNNLOpI}{0.11298\pm 0.00020 {\text( exp.)}}
\newcommand{\CCMOMresultNNLOquality}{436.0/41}
\newcommand{\CCBMOMresultNNNLO}{0.13119\pm 0.00128 {\text( exp.)}\pm 0.0193 {\text( scale)}}
\newcommand{\CCBMOMresultNNNLOpII}{0.0193 {\text( scale)}}
\newcommand{\CCBMOMresultNNNLOpI}{0.13119\pm 0.00128 {\text( exp.)}}
\newcommand{\CCBMOMresultNNNLOquality}{39.7/38}
\newcommand{\CCBMOMresultNNNLOalphazero}{0.88\pm 0.05 {\text( exp.)} \pm 0.305 {(scale)}}
\newcommand{\CCBMOMresultNNNLOMilan}{1.50\pm 0.14 {\text( exp.)}}
\newcommand{\CCBDCpowone}{-4.31\times10^{5}\pm 4.05\times10^{4} {\text( exp.)}}
\newcommand{\CCBMOMresultNNLO}{0.11973\pm 0.00119 {\text( exp.)}\pm 0.0013 {\text( scale)}}
\newcommand{\CCBMOMresultNNLOpII}{0.0013 {\text( scale)}}
\newcommand{\CCBMOMresultNNLOpI}{0.11973\pm 0.00119 {\text( exp.)}}
\newcommand{\CCBMOMresultNNLOquality}{41.3/39}
\newcommand{\CCBMOMresultNNLOalphazero}{0.42\pm 0.02 {\text( exp.)} \pm 0.01 {(scale)}}
\newcommand{\CCBMOMresultNNLOMilan}{1.57\pm 0.29 {\text( exp.)}}
\newcommand{\CCbMOMresultNNNLO}{0.13021\pm 0.00132 {\text( exp.)}\pm 0.0176 {\text( scale)}}
\newcommand{\CCbMOMresultNNNLOpII}{0.0176 {\text( scale)}}
\newcommand{\CCbMOMresultNNNLOpI}{0.13021\pm 0.00132 {\text( exp.)}}
\newcommand{\CCbMOMresultNNNLOquality}{39.7/38}
\newcommand{\CCbMOMresultNNNLOalphazero}{0.86\pm 0.05 {\text( exp.)} \pm 0.285 {(scale)}}
\newcommand{\CCbMOMresultNNNLOMilan}{1.50\pm 0.15 {\text( exp.)}}
\newcommand{\CCbDCpowone}{-4.12\times10^{5}\pm 4.21\times10^{4} {\text( exp.)}}
\newcommand{\CCbMOMresultNNLO}{0.11958\pm 0.00120 {\text( exp.)}\pm 0.0012 {\text( scale)}}
\newcommand{\CCbMOMresultNNLOpII}{0.0012 {\text( scale)}}
\newcommand{\CCbMOMresultNNLOpI}{0.11958\pm 0.00120 {\text( exp.)}}
\newcommand{\CCbMOMresultNNLOquality}{41.2/39}
\newcommand{\CCbMOMresultNNLOalphazero}{0.43\pm 0.02 {\text( exp.)} \pm 0.01 {(scale)}}
\newcommand{\CCbMOMresultNNLOMilan}{1.57\pm 0.29 {\text( exp.)}}
\newcommand{\CCYMOMresultNNNLO}{0.12778\pm 0.00142 {\text( exp.)}\pm 0.0134 {\text( scale)}}
\newcommand{\CCYMOMresultNNNLOpII}{0.0134 {\text( scale)}}
\newcommand{\CCYMOMresultNNNLOpI}{0.12778\pm 0.00142 {\text( exp.)}}
\newcommand{\CCYMOMresultNNNLOquality}{39.6/38}
\newcommand{\CCYMOMresultNNNLOalphazero}{0.81\pm 0.05 {\text( exp.)} \pm 0.24 {(scale)}}
\newcommand{\CCYMOMresultNNNLOMilan}{1.50\pm 0.16 {\text( exp.)}}
\newcommand{\CCYDCpowone}{-3.62\times10^{5}\pm 4.68\times10^{4} {\text( exp.)}}
\newcommand{\CCYMOMresultNNLO}{0.11934\pm 0.00122 {\text( exp.)}\pm 0.0011 {\text( scale)}}
\newcommand{\CCYMOMresultNNLOpII}{0.0011 {\text( scale)}}
\newcommand{\CCYMOMresultNNLOpI}{0.11934\pm 0.00122 {\text( exp.)}}
\newcommand{\CCYMOMresultNNLOquality}{41.0/39}
\newcommand{\CCYMOMresultNNLOalphazero}{0.43\pm 0.02 {\text( exp.)} \pm 0.01 {(scale)}}
\newcommand{\CCYMOMresultNNLOMilan}{1.57\pm 0.29 {\text( exp.)}}

%% file: MOM-def.tex
\newcommand{\eVdist}{\kern-0.06667em}
\newcommand{\GeV}{{\,\text{Ge}\eVdist\text{V\/}}}

\newcommand{\colorfulNNLO}{{CoLoRFulNNLO\xspace}}

\newcommand\prog[1]     {{\tt #1}}

\newcommand{\ABCmanualnew}{
\hline$A^{{\langle (1-T)^{1} \rangle}}_0$     &            2.1034(1) & 2.1034701            & 2.1035               &            2.10344(3) \\
      $B^{{\langle (1-T)^{1} \rangle}}_0$     &           44.995(1)  &                      &           44.999(2)  &           44.99(5)  \\
      $C^{{\langle (1-T)^{1} \rangle}}_0$     &          979.6(6)    &                      &          867(21)     &         1100(30)     \\\hline
\hline$A^{{\langle C^{1} \rangle}}_0$         &            8.6332(5) & 8.6378902            & 8.6379               &            8.6378(1) \\
      $B^{{\langle C^{1} \rangle}}_0$         &          172.834(5)  & 172.85900            &          172.778(7)  &          172.8(3)    \\
      $C^{{\langle C^{1} \rangle}}_0$         &         3525(3)      &                      &         3212(88)     &         4200(100)    \\\hline
\hline$A^{{\langle (1-T)^{2} \rangle}}_0$     &            0.19019(1) & 0.1901961            & 0.1902               &            0.190190(5) \\
      $B^{{\langle (1-T)^{2} \rangle}}_0$     &            6.25943(7) &                      &            6.2595(4) &            6.2568(4) \\
      $C^{{\langle (1-T)^{2} \rangle}}_0$     &          175.17(5)   &                      &          172(1)      &          182.9(2)   \\\hline
\hline$A^{{\langle C^{2} \rangle}}_0$         &            2.4316(1) & 2.4316479            & 2.4317               &            2.43160(4) \\
      $B^{{\langle C^{2} \rangle}}_0$         &           81.1882(9) &                      &           81.184(5)  &           81.160(5)  \\
      $C^{{\langle C^{2} \rangle}}_0$         &         2231.7(6)    &                      &         2220(12)     &         2332(2)      \\\hline
\hline$A^{{\langle (1-T)^{3} \rangle}}_0$     &            0.029875(2) & 0.0298753            & 0.02988              &            0.029874(1) \\
      $B^{{\langle (1-T)^{3} \rangle}}_0$     &            1.12852(1) &                      &            1.1284(1) &            1.1278(1) \\
      $C^{{\langle (1-T)^{3} \rangle}}_0$     &           34.71(1)   &                      &           35.3(2)    &           36.17(3)  \\\hline
\hline$A^{{\langle C^{3} \rangle}}_0$         &            1.07919(7) & 1.0792137            & 1.0792               &            1.07919(3) \\
      $B^{{\langle C^{3} \rangle}}_0$         &           42.7709(4) &                      &           42.771(3)  &           42.752(3)  \\
      $C^{{\langle C^{3} \rangle}}_0$         &         1304.4(3)    &                      &         1296(6)      &         1360.8(8)    \\\hline
\hline$A^{{\langle (1-T)^{4} \rangle}}_0$     &            0.0058583(6) & 0.0058581            & 0.005858             &            0.0058576(3) \\
      $B^{{\langle (1-T)^{4} \rangle}}_0$     &            0.246380(4) &                      &            0.24637(3) &            0.24619(5) \\
      $C^{{\langle (1-T)^{4} \rangle}}_0$     &            8.142(3)  &                      &            8.13(4)   &            8.447(8) \\\hline
\hline$A^{{\langle C^{4} \rangle}}_0$         &            0.56849(4) & 0.5685012            & 0.5685               &            0.56848(2) \\
      $B^{{\langle C^{4} \rangle}}_0$         &           25.8174(3) &                      &           25.816(2)  &           25.804(3)  \\
      $C^{{\langle C^{4} \rangle}}_0$         &          845.4(1)    &                      &          843(3)      &          879.1(5)    \\\hline
\hline$A^{{\langle (1-T)^{5} \rangle}}_0$     &            0.0012948(1) & 0.0012947            & 0.001295             &            0.0012946(9) \\
      $B^{{\langle (1-T)^{5} \rangle}}_0$     &            0.060085(1) &                      &            0.06009(1) &            0.06003(2) \\
      $C^{{\langle (1-T)^{5} \rangle}}_0$     &            2.1170(9) &                      &            2.109(9)  &            2.184(2) \\\hline
\hline$A^{{\langle C^{5} \rangle}}_0$         &            0.32721(3) & 0.3272163            & 0.3272               &            0.32720(1) \\
      $B^{{\langle C^{5} \rangle}}_0$         &           16.8743(2) &                      &           16.873(1)  &           16.865(3)  \\
      $C^{{\langle C^{5} \rangle}}_0$         &          586.9(1)    &                      &          585(3)      &          607.4(4)    \\\hline

}

%% file: MOM-abs.tex
We consider a method for determining the QCD strong coupling constant 
using fits of perturbative predictions for event shape  averages  to data 
collected at the LEP, PETRA, PEP and TRISTAN colliders. To obtain highest 
accuracy predictions we use a combination of perturbative ${\cal{O}}(\alpha_{S}^{3})$ 
calculations and estimations of the ${\cal{O}}(\alpha_{S}^{4})$ perturbative 
coefficients from data. We account for non-perturbative effects using modern 
Monte Carlo event generators and analytic hadronization models. 
 The obtained results show that the total precision of the $\alpha_{S}$ 
determination cannot be improved significantly with the higher order perturbative QCD
corrections alone, but primarily  requires a deeper understanding of the non-perturbative effects.

%% file: MOM-txt.tex
\newpage
%%%%%%%%%%%%%%%%%%%%%%%%%%%%%%%%%%%%%%%%%%%%%%%%%%%%%%%%%%%%%%%%%%%%%%%%
\section{Introduction}
\label{sec:introduction}

Measurements using hadronic final states in $e^{+}e^{-}$ annihilation 
have provided detailed experimental tests of Quantum Chromodynamics (QCD), 
the theory of the strong interaction in the Standard Model. These
measurements were based on comparisons of moments and differential 
distributions of event shapes or jet rates to perturbative predictions. 
As new data are not foreseen in the near future, the progress in such 
measurements depends wholly on improvements in the theoretical (and phenomenological) description 
of these observables.
  Moreover, in multiple QCD analyses of the LEP data in the past, it was shown that the experimental uncertainties play 
a relatively small role in comparison to the theory-related uncertainties.   

 This situation rises some important questions. First of all, would the increasingly precise 
perturbative QCD (pQCD) calculations amended with resummation techniques be able to improve the precision of the results in QCD 
studies without any new data? And if not, what would be the limiting factors for 
the precision of QCD studies in the future and what should be done to eliminate them? 
To answer these questions we perform a QCD analysis with state-of-the-art pQCD calculations extended 
with estimations of higher-order corrections not known at present.

Fully differential  pQCD  calculations for the production of three partonic 
jets in $e^{+}e^{-}$ hadronic annihilation are available to ${\cal O}(\alpha_{S}^{3})$
accuracy~\cite{GehrmannDeRidder:2007hr,GehrmannDeRidder:2009dp,Weinzierl:2009ms,
Weinzierl:2009yz,DelDuca:2016csb,DelDuca:2016ily}, 
which corresponds to next-to-next-to-leading order (NNLO) in QCD perturbation 
theory for this process. Four- and five-jet production~\cite{Signer:1996bf,
Dixon:1997th,Nagy:1997yn,Nagy:1998bb,Campbell:1998nn,Weinzierl:1999yf}, 
as well as the total cross section~\cite{Baikov:2012zn} 
are also known including terms at ${\cal O}(\alpha_{S}^{3})$,\footnote{In the 
case of four- and five-jet production, ${\cal O}(\alpha_{S}^{3})$ accuracy 
corresponds to next-to-leading order (NLO) and leading order (LO) in 
perturbative QCD. However, NLO corrections to five-jet production~\cite{Frederix:2010ne} 
(and up to seven-jet production in the leading color approximation~\cite{Becker:2011vg}) 
are also known.} therefore it is possible to make predictions for any infrared-safe 
observable at this level of accuracy. Although higher-order corrections are presently 
not known, it is in principle possible to estimate such corrections from data and 
therefore to obtain ``predictions'' at ${\cal O}(\alpha_{S}^{4})$.
This approach is obviously limited to cases of observables for which only a 
small number of coefficients of the perturbative expansion should be estimated, 
such as event shape moments. In this paper we present an implementation of 
this approach with the aim of assessing the impact of these terms on 
possible future extractions of the strong coupling with exact predictions at 
${\cal O}(\alpha_{S}^{4})$.

When confronting calculations based on QCD perturbation theory 
(of any order) with data, it must be kept in mind that although in 
$e^+e^-$ annihilation strong interactions occur only in the final state, 
nevertheless, the observed quantities are affected by hadronization and power 
corrections. These corrections must either be extracted from Monte Carlo 
predictions or computed using analytic models. Below, we consider both of 
these approaches for describing non-perturbative effects and perform 
simultaneous fits of $\alpha_{S}(M_Z)$ and the ${\cal O}(\alpha_{S}^{4})$ 
perturbative coefficients to event shape moments (together with model 
parameters for analytic hadronization models) for thrust\footnote{More 
precisely, we consider the quantity $\tau \equiv 1-T$, where $T$ is the thrust.}~\cite{Brandt:1964sa,
Farhi:1977sg} and the $C$-parameter~\cite{Parisi:1978eg,Donoghue:1979vi}.

 Anticipating some of our findings, we observe a clear discrepancy between
results obtained with fits using Monte Carlo and analytic hadronization models, even
after the inclusion of higher orders in pQCD and extending the analytic models to
${\cal O}(\alpha_{S}^{4})$. This implies that efforts to significantly improve the overall
precision of $\alpha_{S}$ extractions in the future must address not only the computation
of higher-order pQCD corrections, but also the refined modeling of non-perturbative effects.

 We expect that the presented analysis will provide valuable input for the 
planing of $\alpha_{S}(M_Z)$ measurements and data taking at future $e^{+}e^{-}$ facilities.

\section{Theory predictions}
\label{sec:theory}

The $n$-th moment of an event shape variable $O$ is defined by
$$
\langle O^{n} \rangle = \frac{1}{\sigma_{\mathrm{tot}}} 
\int_{O_{\mathrm{min}}}^{O_{\mathrm{max}}} O^n \frac{\mathrm{d}\sigma(O)}{\mathrm{d}O}\mathrm{d}O,
$$
where $\sigma_{\mathrm{tot}}$ stands for the total hadronic cross section and 
$[O_{\mathrm{min}},O_{\mathrm{max}}]$ is the kinematically allowed range of the 
observable $O$.

The fixed-order prediction for the $n$-th moment of $O$ at 
a reference renormalization scale $\mu = \mu_0$, normalized to the LO cross 
section $\sigma_0$ for $e^+e^- \to \mbox{hadrons}$ reads:
\begin{equation*}
\begin{split}
\frac{1}{\sigma_0} 
\int_{O_{\mathrm{min}}}^{O_{\mathrm{max}}} O^n \frac{\mathrm{d}\sigma(O)}{\mathrm{d}O}
\mathrm{d}O
&=\frac{\alpha_{S}(\mu_0)}{2\pi}A^{{\langle O^{n} \rangle}}_0
+\left(\frac{\alpha_{S}(\mu_0)}{2\pi}\right)^2 B^{{\langle O^{n} \rangle}}_0
\epjconly{\\}
+\left(\frac{\alpha_{S}(\mu_0)}{2\pi}\right)^3 C^{{\langle O^{n} \rangle}}_0
\draftonly{\\}
\arxivonly{\\}&
+\left(\frac{\alpha_{S}(\mu_0)}{2\pi}\right)^4 D^{{\langle O^{n} \rangle}}_0
+{\cal O}(\alpha_{S}^5). 
\end{split}
\end{equation*}
Throughout the paper we employ the $\overline{\mathrm{MS}}$ renormalization scheme 
and $\alpha_{S}$ (without a superscript) always denotes the strong coupling in this 
scheme.
The coefficients $A^{\langle O^{n} \rangle}_0$, $B^{\langle O^{n} \rangle}_0$ 
and $C^{\langle O^{n} \rangle}_0$ for moments of standard event shapes have been 
known for some time~\cite{GehrmannDeRidder:2009dp,Weinzierl:2009yz}.
In this paper, we use the \colorfulNNLO~\cite{Somogyi:2006da,Somogyi:2006db,DelDuca:2016ily} 
approach to recompute these coefficients with high numerical precision, 
see Tab.~\ref{tab:PT-data}. 

\begin{table}\centering\small
\begin{tabular}{|c|c|c|c|c|}\hline
 Coefficient & This work & Analytic & Ref.~\cite{GehrmannDeRidder:2009dp} & Ref.~\cite{Weinzierl:2009yz} 
 \\\hline\hline
 \ABCmanualnew
\end{tabular}
\caption{\label{tab:PT-data} LO, NLO and NNLO contributions to the moments 
 of event shapes. For the details on the analytic calculation see App.~\protect\ref{app:AB}.}
\end{table}

This allows us to extend the extraction of 
the strong coupling constant from these observables to N$^{3}$LO with a 
simultaneous extraction of the perturbative coefficients $D^{\langle O^{n} \rangle}_0$ from 
the data. 

However, the experimentally measured  event shape moments are normalized to the total 
hadronic cross section, and the perturbative expansion of $\langle O^{n} \rangle$ 
is given by
\begin{equation*}
\begin{split}
\langle O^{n} \rangle
&=\frac{\alpha_{S}(\mu_0)}{2\pi}\bar{A}^{\langle O^{n} \rangle}_0
+\left(\frac{\alpha_{S}(\mu_0)}{2\pi}\right)^2 \bar{B}^{\langle O^{n} \rangle}_0
\epjconly{\\&}
+\left(\frac{\alpha_{S}(\mu_0)}{2\pi}\right)^3 \bar{C}^{\langle O^{n} \rangle}_0
+\left(\frac{\alpha_{S}(\mu_0)}{2\pi}\right)^4 \bar{D}^{\langle O^{n} \rangle}_0
\draftonly{\\&}
\arxivonly{\\&}
+{\cal O}(\alpha_{S}^5). 
\end{split}
\end{equation*}
The relations between $\bar{A}^{\langle O^{n} \rangle}_0$, 
$\bar{B}^{\langle O^{n} \rangle}_0$, $\bar{C}^{\langle O^{n} \rangle}_0$, 
$\bar{D}^{\langle O^{n} \rangle}_0$ and $A^{\langle O^{n} \rangle}_0$, 
$B^{\langle O^{n} \rangle}_0$, $C^{\langle O^{n} \rangle}_0$, 
$D^{\langle O^{n} \rangle}_0$ are straightforward to obtain using
\begin{equation*}
\begin{split}
\frac{\sigma_0}{\sigma_{\mathrm{tot}}} &= 
    1 - \frac{\alpha_{S}}{2\pi} A_{\mathrm{tot}}
    + \left(\frac{\alpha_{S}}{2\pi}\right)^2 
    \left(A_{\mathrm{tot}}^2 - B_{\mathrm{tot}}\right)\epjconly{\\&}
    - \left(\frac{\alpha_{S}}{2\pi}\right)^3 
    \left(A_{\mathrm{tot}}^3 - 2A_{\mathrm{tot}}B_{\mathrm{tot}} + C_{\mathrm{tot}}\right)
    + \mathcal{O}(\alpha_{S}^4)\,,
\end{split}
\end{equation*}
and we find
\begin{align}
\bar{A}^{\langle O^{n} \rangle}_0 &= A^{\langle O^{n} \rangle}_0\,,
\notag
\\
\bar{B}^{\langle O^{n} \rangle}_0 &= B^{\langle O^{n} \rangle}_0 - A_{\mathrm{tot}} A^{\langle O^{n} \rangle}_0\,,
\notag
\\
\bar{C}^{\langle O^{n} \rangle}_0 &= C^{\langle O^{n} \rangle}_0 - A_{\mathrm{tot}} B^{\langle O^{n} \rangle}_0
    + \left(A_{\mathrm{tot}}^2 - B_{\mathrm{tot}}\right) A^{\langle O^{n} \rangle}_0\,,
\notag
\\
\bar{D}^{\langle O^{n} \rangle}_0 &= D^{\langle O^{n} \rangle}_0 - A_{\mathrm{tot}} C^{\langle O^{n} \rangle}_0
    + \left(A_{\mathrm{tot}}^2 - B_{\mathrm{tot}}\right) B^{\langle O^{n} \rangle}_0\epjconly{\notag\\&}
    - \left(A_{\mathrm{tot}}^3 - 2A_{\mathrm{tot}}B_{\mathrm{tot}} + C_{\mathrm{tot}}\right)
    A^{\langle O^{n} \rangle}_0.
\notag
\end{align}
The coefficients $A_{\mathrm{tot}}$,  $B_{\mathrm{tot}}$ and  $C_{\mathrm{tot}}$ are listed 
in App.~\ref{app:ABCtot}.

Finally, to perform a simultaneous fit of multiple data points at different 
center-of-mass energies, we use the four-loop running of $\alpha_{S}(\mu)$ 
\begin{align}
\mu^2\frac{d}{d \mu^2} \frac{\alpha_S(\mu)}{4\pi} = - \left(\frac{\alpha_S(\mu)}{4\pi}\right)^2 \sum_{n} \beta_n \left(\frac{\alpha_S(\mu)}{4\pi}\right)^n\,,
\end{align}
 with $\beta_0 = (11 C_A - 2 N_F)/3$, 
$\beta_1 = (34 C_A^2 - 10 C_A N_F - 6 C_F N_F)/3$ and 
$\beta_2 = (2857 C_A^3 - 1415 C_A^2 N_F - 615 C_A C_F N_F + 54 C_F^2 N_F 
+ 79 C_A N_F^2 +\epjconly{$ \\ $+} 66 C_F N_F^2)/54$. We are using the customary normalization of 
$T_R = 1/2$ for the color charge operators, thus in QCD we have $C_A = N_c = 3$ 
and $C_F = (N_c^2-1)/(2N_c) = 4/3$, while $N_F$ denotes the number of light quark 
flavors.
The corresponding dependence of the perturbative coefficients $\bar{A}^{\langle O^{n} \rangle}$, 
$\bar{B}^{\langle O^{n} \rangle}$, $\bar{C}^{\langle O^{n} \rangle}$ and $\bar{D}^{(n)}$ on scale
 read:
\begin{align}
\bar{A}^{\langle O^{n} \rangle} &= \bar{A}^{\langle O^{n} \rangle}_0\,,
\notag
\\
\bar{B}^{\langle O^{n} \rangle} &= \bar{B}_0^{\langle O^{n} \rangle} + \frac12 \bar{A}^{\langle O^{n} \rangle}_0 \beta_0 L\,,
\notag
\\
\bar{C}^{\langle O^{n} \rangle} &= \bar{C}_0^{\langle O^{n} \rangle} + \bar{B}^{\langle O^{n} \rangle}_0 \beta_0 L 
    + \frac14 \bar{A}^{\langle O^{n} \rangle}_0(\beta_1 + \beta_0^2 L)L\,,
\notag
\\
\bar{D}^{\langle O^{n} \rangle} &= \bar{D}_0^{\langle O^{n} \rangle} + \frac32 \bar{C}_0^{\langle O^{n} \rangle}\beta_{0}L 
    + \frac12 \bar{B}_0^{\langle O^{n} \rangle}\left(\beta_{1}+\frac32 \beta_{0}^2L\right)L
\\ &
    + \frac18 \bar{A}_0^{\langle O^{n} \rangle}\left(\beta_{2}+\frac52\beta_{1}\beta_{0}L+\beta_{0}^3L^2\right)L,
\notag
\end{align}
where we have $L=\ln(\mu^{2}/\mu_{0}^{2})$.

We take into account the effect of non-vanishing $b$-quark mass on 
the predictions for the $A^{\langle O^{n} \rangle}$ and $B^{\langle O^{n} \rangle}$ 
coefficient by subtracting the fraction of $b$-quark events, $r_b(Q)$, from the massless 
result and adding back the corresponding massive prediction obtained with the 
\prog{Zbb4}~\cite{Nason:1997tz} program,
\begin{align}
A^{\langle O^{n} \rangle} &= A^{\langle O^{n} \rangle}_{m_b=0}(1-r_b(Q)) + r_b(Q) A^{\langle O^{n} \rangle}_{m_b\ne 0}\,,
\notag
\\
B^{\langle O^{n} \rangle} &= B^{\langle O^{n} \rangle}_{m_b=0}(1-r_b(Q)) + r_b(Q) B^{\langle O^{n} \rangle}_{m_b\ne 0}\,.
\notag
\end{align}
\section{Data sets}
\label{sec:data}
For the performed analysis, one minus thrust $\tau\equiv(1-T)$ and the $C$-parameter 
($C$) were selected. The selection of these particular observables is motivated 
by the abundance of available measurements.
More specifically, in this analysis we considered data sets from the 
ALEPH, AMY, DELPHI, HRS, JADE, L3, MARK, MARKII, OPAL and TASSO 
experiments, see Tab.~\ref{tab:data} for details.

 As the theory predictions for all the measured event shape moments were calculated in the \colorfulNNLO~  \epjconly{\\} framework and 
 the hadronization corrections can be obtained consistently for all the event shape moments, the most complete analysis of the data might 
 include simultaneously all the measured moments.
 
However, as most of the moments of the event shapes are quite strongly correlated~\cite{Pahl:2007zz},
a simultaneous analysis of all available data would require taking these 
correlations into account.  Unfortunately, very few data sets contain this crucial information and
therefore, we have limited our analysis only to 
the first moments,  i.e.\ averages of the event shapes.
From the two available sets of measurements with the same data in the range $\sqrt{s}=133-183\GeV$, 
available from Ref.~\cite{Reinhardt:2001zz} and Ref.~\cite{Abreu:1999rc} 
the measurements from Ref.~\cite{Abreu:1999rc} were used in the analysis.

\begin{table}\centering\small
\begin{tabular}{|c|c|c|c|c|c|}\hline
             & \multicolumn{2}{c|} {Measured   } & \multicolumn{2}{c|} {Used    }     \\\hline
\epjconly{
             &   Obser-      &  Points,                            &  Obser-      & Points,                           \\
Source       &   vables      &   $\sqrt{s}$ range                  &  vables      & $\sqrt{s}$ range \\
             &               &   ($\GeV)$                          &              & ($\GeV)$                         \\\hline\hline
}

\draftonly{
             &                &  Points,                           &   & Points,                           \\
Source       &   Observables  &   $\sqrt{s}$ range                 &  Observables & $\sqrt{s}$ range \\
             &                &   ($\GeV)$                         &              & ($\GeV)$                         \\\hline\hline
}

\arxivonly{
             &                &  Points,                           &   & Points,                           \\
Source       &   Observables  &   $\sqrt{s}$ range                 &  Observables & $\sqrt{s}$ range \\
             &                &   ($\GeV)$                         &              & ($\GeV)$                         \\\hline\hline
}

\MOMtabularinputoneminusT\hline\hline
\MOMtabularinputC\hline
\end{tabular}
\caption{Available measurements and  data used in the analysis. Here we use $\tau\equiv(1-T)$.}    
\label{tab:data}
\end{table}

\section{Modeling of non-perturbative corrections}
\label{sec:modeling}
As discussed in the Introduction, the modeling of non-perturbative 
corrections is essential in order to perform a meaningful comparison 
of theoretical predictions with data. One option for obtaining the 
hadronization corrections is to extract them from Monte Carlo simulations.  
Recent examples of this approach include the studies of the energy-energy 
correlation~\cite{Kardos:2018kqj} and the two-jet rate~\cite{Verbytskyi:2019zhh}.

Some previous extractions of the strong coupling from event shape 
moments~\cite{Gehrmann:2009eh,Pahl:2007zz} and event shape distributions~\cite{Gehrmann:2012sc} have used an analytic hadronization model 
based on the dispersive approach to power corrections~\cite{Dokshitzer:1995qm,
Dokshitzer:1997ew,Dokshitzer:1998pt}. An important ingredient of 
this model is the relation between the strong coupling defined in the 
$\overline{\mathrm{MS}}$ scheme and the effective soft coupling 
$\alpha_{S}^{CMW}$ in the Catani--Marchesini--Webber (CMW) scheme. As the extension of 
$\alpha_{S}^{CMW}$ beyond NLL accuracy is believed not to be unique~\cite{Banfi:2018mcq,
Catani:2019rvy}, the coefficients in this relation are ``scheme-dependent''. 
However, in one particular proposal, the relation between $\alpha_{S}$ 
and $\alpha_{S}^{CMW}$ has recently been computed up to $\mathcal{O}(\alpha_{S}^{4})$ 
accuracy~\cite{Catani:2019rvy}, which allows us to implement a consistent analytic 
model of hadronization corrections at this order in the perturbative expansion.

Below, we pursue both options and use Monte Carlo tools as well as the 
analytic approach to estimate the non-perturbative corrections.

\subsection{Monte Carlo hadronization models}
\label{sec:mc}
In this work we use 
the Monte Carlo event generation setups  similar to those in
previous comparable studies~\cite{Kardos:2018kqj}. 
We made use of the \prog{Herwig7.2.0}~\cite{Bellm:2015jjp} and  
\prog{Sherpa2.2.8}~\cite{Gleisberg:2008ta} Monte Carlo event generators (MCEGs) 
with similar setups for perturbative calculations, but different hadronization 
models. 
The QCD matrix elements for the $e^+e^- \to Z/\gamma \to 2,3,4,5$ parton 
processes were generated using \prog{MadGraph5}~\cite{Alwall:2011uj} and the 
 \prog{OpenLoops}~\cite{Cascioli:2011va} one-loop library. The 2-parton final state 
process was computed at NLO accuracy in perturbative QCD.
The generated events were hadronized using the Lund hadronization 
model~\cite{Andersson:1983ia} or the cluster hadronization model~\cite{Webber:1983if}.
In the following, the setup labelled as $H^L$ denotes predictions computed 
with \prog{Herwig7.2.0} employing the Lund hadronization model~\cite{Andersson:1983ia}, 
$H^C$ denotes \prog{Herwig7.2.0} predictions obtained with the cluster hadronization 
model~\cite{Webber:1983if}, and finally $S^C$ denotes results obtained using  
\prog{Sherpa2.2.8} with the cluster hadronization model~\cite{Winter:2003tt}.

For the study, predictions of event shape moments were calculated 
from MC generated events at hadron  and parton levels
($ \langle O^{n} \rangle _{\mathrm{MC\ hadrons}}$ and  
($ \langle O^{n} \rangle _{\mathrm{MC\ partons}}$).
To take into account that the presence of a shower cut-off 
$Q_0\approx{\cal O}(1\GeV)$ in Monte Carlo programs affects the event 
shape distributions (e.g.\ see Refs.~\cite{Hoang:2018zrp,Baumeister:2020mpm}) both parton 
and hadron level MC predictions were calculated with 
several different values of the parton shower cut-offs and 
extrapolated to $Q_0\rightarrow 0\GeV$.
\begin{figure}[htbp]\centering
\includegraphics[width=\BROADFIGWIDTH]{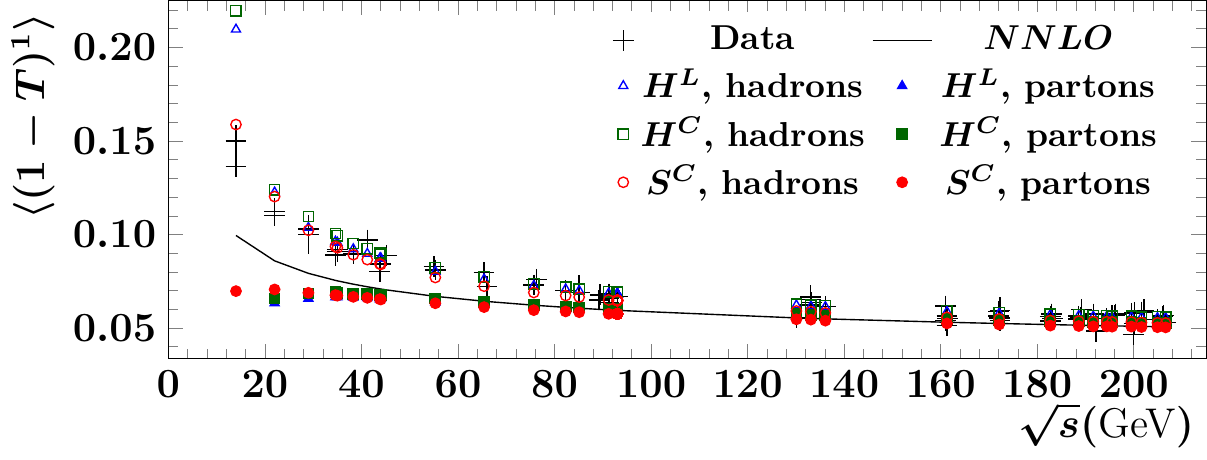}
\includegraphics[width=\BROADFIGWIDTH]{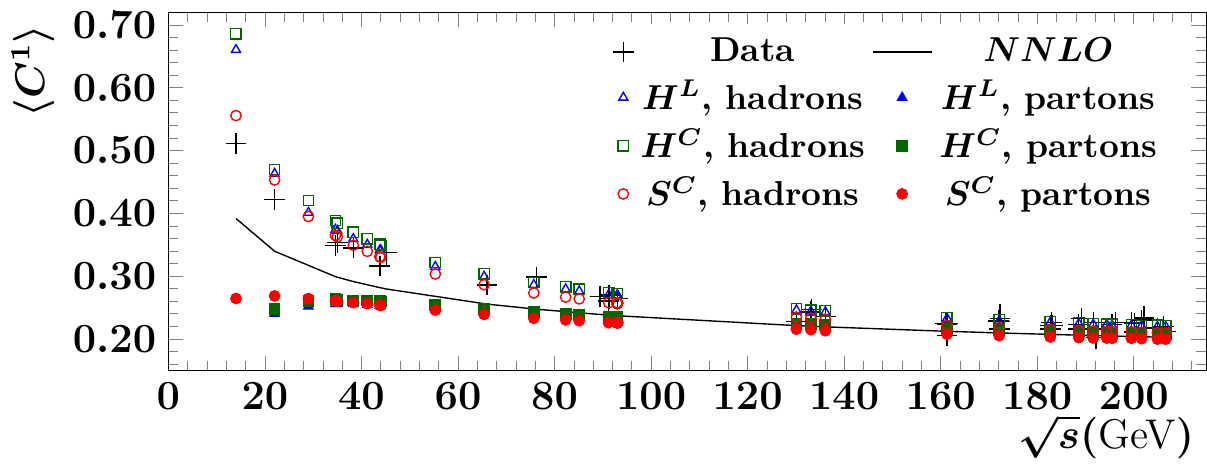}\\
\caption{Data and predictions by MCEGs extrapolated to $Q_0=0\GeV$. 
The NNLO result was computed using $\alpha_{S}(M_Z) = 0.118$.}
\label{fig:mc}
\end{figure}
Fig.~\ref{fig:mc} shows the final results obtained with the various MCEG setups 
after extrapolation to $Q_0 = 0\GeV$, together with the experimental measurements. 
The hadron and parton level MC predictions seen in Fig.~\ref{fig:mc} provide reasonable 
descriptions of the data as well as the NNLO perturbative results for a wide range of 
center-of-mass energies. However, the MC predictions at lowest $\sqrt{s}$ show non-physical 
behavior, i.e.\ $\langle O^{n} \rangle$ increases with $\sqrt{s}$ for the parton 
level results. 
In order to analyze only the data that can be adequately described by the Monte Carlo 
modeling, we exclude measurements with $\sqrt{s}<29\GeV$ from the analysis.
 In practice this criterion is much weaker 
than a requirement that MC matches the data well or that the 
subleading power corrections to the analytic hadronization models are small.
However, this choice, in addition to retaining as much of the data as is reasonable,
serves to highlight the discrepancies between the Monte Carlo and analytic hadronization
models in regions where hadronization effects are most pronounced, i.e.\ at low energies.

\begin{figure}[htbp]\centering
\includegraphics[width=\BROADFIGWIDTH]{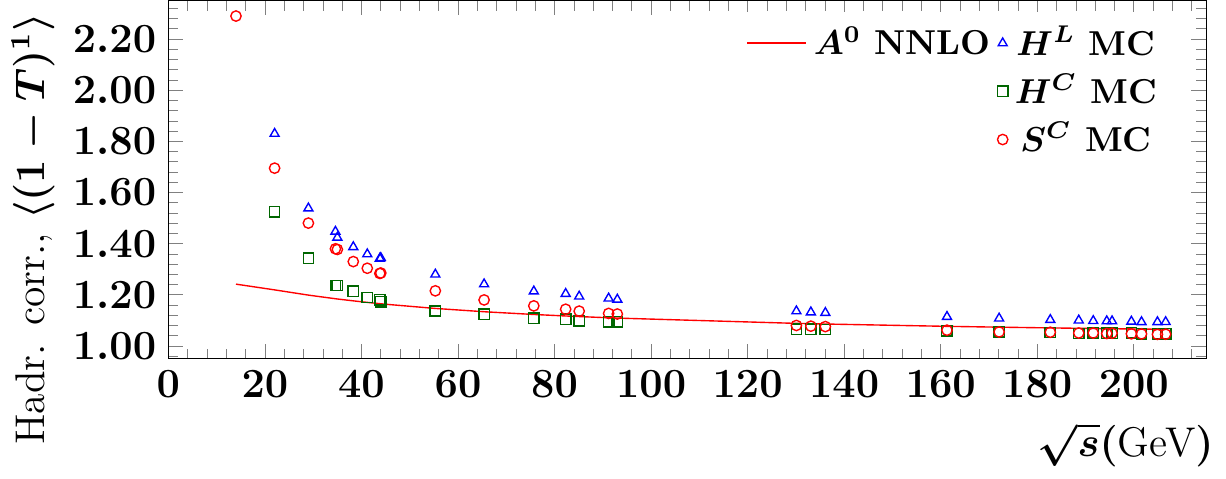}\\
\includegraphics[width=\BROADFIGWIDTH]{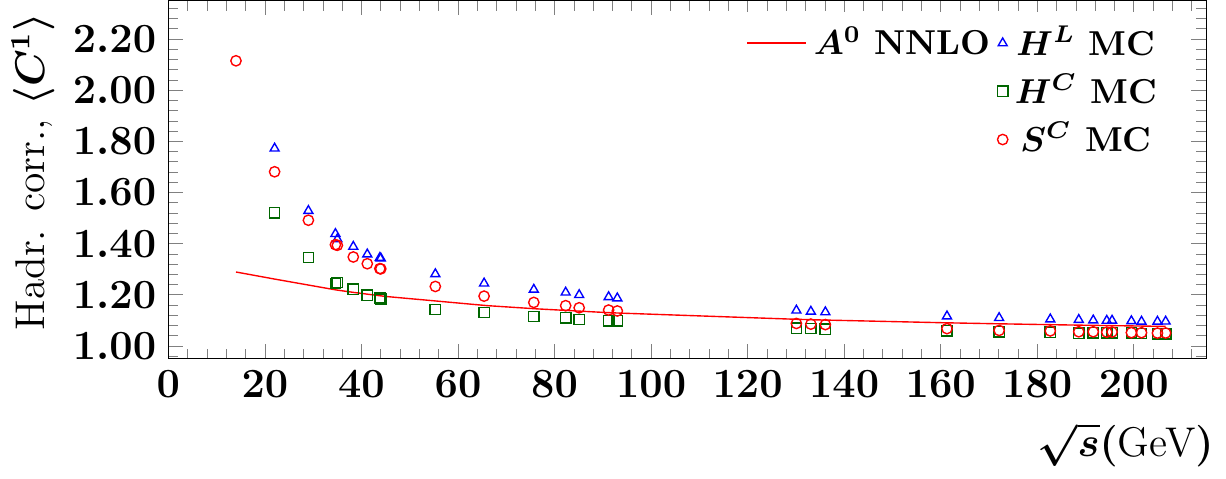}\\
\caption{Hadronization corrections extracted from MC  generated samples after 
extrapolation to $Q_0=0\GeV$ and the hadronization corrections from the ${\cal A}_{0}$ 
scheme calculated as ratios of the hadron and parton level predictions using 
$\alpha_S(M_Z)=0.118$, $\alpha_0(\mu_I)=0.5$ and $\mathcal{M}=1.49$. The hadronization 
corrections from the ${\cal A}_{T}$- and cusp-schemes are not shown, but 
these very closely follow the hadronization corrections from the ${\cal A}_{0}$-scheme.
}
\label{fig:had}
\end{figure}

Finally, the correction of theory predictions for hadronization was 
implemented in the analysis as follows,
$$ 
\langle O^{n} \rangle _{\mathrm{corrected}} 
= \langle O^{n} \rangle _{\mathrm{theory}}\times
\frac{ \langle O^{n} \rangle _{\mathrm{MC\ hadrons},\,Q_0=0\GeV}}{ \langle O^{n} 
\rangle _{\mathrm{MC\ partons},\,Q_0=0\GeV}}\,.
$$
The hadronization correction factors for different center-of-mass
energies, observables and MC setups are shown in Fig.~\ref{fig:had}. 
\FloatBarrier
\subsection{Analytic hadronization models}
\label{sec:analytic}
The dispersive model of analytic hadronization corrections 
for event shapes in $e^{+}e^{-}$ annihilation has been worked out in detail in 
Refs.~\cite{Dokshitzer:1995qm,Dokshitzer:1997ew,Dokshitzer:1998pt}. In this model, 
hadronization corrections simply shift the perturbative event shape distributions,
\begin{equation}
\frac{\mathrm{d}\sigma_{\mathrm{hadrons}}(O)}{\mathrm{d}O} 
    = \frac{\mathrm{d}\sigma_{\mathrm{partons}}(O - a_O {\mathcal P})}{\mathrm{d}O}, 
\label{eq:dO-hadr}
\end{equation}
where the power correction ${\mathcal P}$ is universal for all event shapes, while 
the $a_O$ are specific, known constants, e.g.\  $a_{1-T}=2$ and $a_{C}=3\pi$ for $1-T$ 
and the $C$-parameter. Inserting eq.~(\ref{eq:dO-hadr}) into the definition of the 
moments, one obtains the non-perturbative predictions for event shape moments. 
In particular, the effect of hadronization corrections on the  average is additive,
$$
\langle O^1 \rangle_{\mathrm{hadrons}}
    = \langle O^1 \rangle_{\mathrm{partons}} + a_O \mathcal{P}\,,
$$
where $\langle O^1 \rangle_{\mathrm{partons}}$ is the value obtained in fixed-order 
perturbation theory as described in Sect.~\ref{sec:theory}.
 Some deviations from this simple model were discussed very recently
in Ref.~\cite{Luisoni:2020efy}.

Finally, we must compute the power correction ${\cal{P}}$ at $\mathcal{O}(\alpha_{S}^4)$ 
accuracy. The perturbative ingredients of this calculation are the running of the strong 
coupling in the $\overline{\mathrm{MS}}$ scheme and the relation between the coupling 
defined in the $\overline{\mathrm{MS}}$ and the CMW schemes. This relation takes the 
following generic form
\begin{align}
\alpha_{S}^{CMW} = \alpha_{S} \left[1 
    + \frac{\alpha_{S}}{2\pi} K 
    \right.&+ \left(\frac{\alpha_{S}}{2\pi}\right)^{2} L \epjconly{\nonumber\\&}\left.
    + \left(\frac{\alpha_{S}}{2\pi}\right)^{3} M 
    + \mathcal{O}(\alpha_{S}^4)\right]\,.
\label{eq:alpha_CMW}
\end{align}
The value of the $K$ coefficient has been known to coincide with the one-loop 
cusp anomalous dimension for a long time and hence it may be tempting to assume 
that the  cusp anomalous dimension provides a sensible definition of the CMW coupling 
also beyond $\mathcal{O}(\alpha_{S}^2)$. However, this assumption turns out to be 
incorrect\footnote{A simple way to see that the equivalence between the coefficients 
in eq.~(\ref{eq:alpha_CMW}) and the cusp anomalous dimensions cannot hold in general 
is to realise that the latter depend on the factorization scheme of collinear singularities 
while the former should not.} and as mentioned above, it is believed that there is 
no unique extension of $\alpha_{S}^{CMW}$ beyond NLL accuracy. Nevertheless, recently 
several proposals have been made for the definition of the effective soft coupling 
in the literature~\cite{Banfi:2018mcq,Catani:2019rvy}. In particular, Ref.~\cite{Catani:2019rvy} 
introduces the effective soft-gluon coupling ${\mathcal{A}}_{i}^{CMW}$ as
$$
{\mathcal{A}}_{i}^{CMW}(\alpha_{S}) = C_i \frac{\alpha_{S}^{CMW}}{\pi} 
    = C_i \frac{\alpha_{S}}{\pi}\left(1 + \frac{\alpha_{S}}{2\pi} K + \ldots\right)\,,
$$
(here $i$ denotes the type of radiating parton, so $C_q = C_F$ and $C_g = C_A$) 
and proposes two different prescriptions for defining this coupling beyond 
NLL accuracy, denoted by $\mathcal{A}_{0,i}$ and $\mathcal{A}_{T,i}$.\footnote{The 
definition of the soft coupling proposed in Ref.~\cite{Banfi:2018mcq} is equivalent to 
$\mathcal{A}_{T,i}$ of Ref.~\cite{Catani:2019rvy}.} 
We will refer to these cases as the ``$\mathcal{A}_{0}$-scheme'' and the 
``$\mathcal{A}_{T}$-scheme'' below. We note that the complete expression for the $M$ 
coefficient is currently not known in the $\mathcal{A}_{T}$-scheme, hence in our 
analysis we approximate this coefficient with its value in the $\mathcal{A}_{0}$-scheme 
and set $M_T = M_0$. In order to facilitate the comparison of our results with previous 
work~\cite{Gehrmann:2009eh}, we also define the ``cusp-scheme'', in which we simply set 
the $K$, $L$ and $M$ coefficients of eq.~(\ref{eq:alpha_CMW}) equal to the appropriate 
cusp anomalous dimension. In the following, we will denote the results obtained in the 
$\mathcal{A}_{0}$-scheme by $A^0$, in the $\mathcal{A}_{T}$-scheme by $A^T$ and in the 
cusp-scheme by $A^{\mathrm{cusp}}$. The explicit expressions for the $K$, $L$ and $M$ coefficients 
in all three schemes are presented in App.~\ref{app:KLM}.

Finally, the power correction takes the following form up to N${}^3$LO, 
\begin{align}
{\cal{P}}(\alpha_{S},Q,\alpha_{0})&=\frac{4C_F}{\pi^2}\,
\mathcal{M} \times\frac{\mu_I}{Q}\times \,\bigg\lbrace\alpha_0(\mu_I)-\bigg[
                                   \alpha_{S}(\mu_R)  \epjconly{\nonumber \\ &}  + 
                                   \left(K + \beta_0 \left(1 + \ln\frac{\mu_R}{\mu_I}\right)\right) \frac{\alpha_{S}^2(\mu_R)}{2 \pi}
\nonumber \\ &                                   
     + \left(2 L + \left(4 \beta_0 \left(\beta_0 + K\right) + \beta_1\right) \left(1 + \ln\frac{\mu_R}{\mu_I}\right) 
       \right.\epjconly{\nonumber \\ &}\left. + 
        2 \beta_0^2 \ln^2\frac{\mu_R}{\mu_I}\right) \frac{\alpha_{S}^3(\mu_R)}{8 \pi^2}
\nonumber \\ &                                           
     + \left(4 M +  \big(2 \beta_0  \left(12 \beta_0 (\beta_0 + K) + 5 \beta_1\right) + \beta_2   
       \right.\epjconly{\nonumber \\ &}\left.  
          + 4 \beta_1 K + 12 \beta_0 L \big) \left(1 + \ln\frac{\mu_R}{\mu_I}\right)\right. 
\nonumber \\ & 
     +   \beta_0 (12 \beta_0 (\beta_0 + K) + 5 \beta_1) \ln^2\frac{\mu_R}{\mu_I}  \epjconly{\nonumber \\ &}\left. 
     +   4 \beta_0^3 \ln^3\frac{\mu_R}{\mu_I} \right) \frac{\alpha_{S}^4(\mu_R)}{32 \pi^3}
\bigg]
\bigg\rbrace,
\label{eq:P}
\end{align}
where $\mu_I$ is the scale at which the perturbative and non-perturbative couplings 
are matched in the dispersive model and 
$\mathcal{M}$ is the so-called Milan factor~\cite{Dokshitzer:1998pt} with 
an estimated value of $\mathcal{M}_{\rm est.}\pm\delta\mathcal{M}_{\rm est.}=1.49\pm0.30$~\cite{Dokshitzer:1998pt,Gehrmann:2009eh}. 
$\alpha_0(\mu_I)$ 
corresponds to the first moment of the effective coupling below the scale $\mu_I$
$$
\alpha_0(\mu_I) = \frac{1}{\mu_I}\int_0^{\mu_I} {\mathrm d}\mu\, \alpha^{CMW}(\mu)\,,
$$
and it is a non-perturbative parameter of the model. Following the usual choice, we 
will set $\mu_I = 2\GeV$. We note that the value of 
$\alpha_0(\mu_I)$ is in principle scheme-dependent, i.e.\ it depends on the precise 
relation between the strong coupling in the $\overline{\mathrm{MS}}$ and CMW schemes. 
In contrast, $\alpha_{S}(\mu_R)$ always refers to the value of the strong coupling in the 
$\overline{\mathrm{MS}}$ scheme, at scale $\mu_R$.

Before moving on, let us make two comments regarding eq.~(\ref{eq:P}). First, it can 
be argued (see e.g.\ the discussion in ref.~\cite{Gehrmann:2012sc}) that the $L$ and 
$M$ coefficients that appear in ${\cal{P}}(\alpha_{S},Q,\alpha_{0})$ do not coincide 
precisely with those in eq.~(\ref{eq:alpha_CMW}), but receive modifications due to the 
non-inclusive nature of the observables we are studying. As such, $L$ and $M$ in 
eq.~(\ref{eq:P}) might be different for different observables. Second, beyond NLO, 
there is some doubt that the non-inclusive corrections parametrized by the Milan factor 
$\mathcal{M}$ can still be captured by a simple overall multiplicative factor in 
${\cal{P}}(\alpha_{S},Q,\alpha_{0})$. Nevertheless, here we take the pragmatic viewpoint 
that eq.~(\ref{eq:P}) provides a reasonable model for non-perturbative corrections. 
In particular, as in other analyses \cite{Gehrmann:2009eh,Gehrmann:2012sc}, the Milan 
factor appears as a multiplicative constant. However, the applicability of this approach 
should be investigated in detail in future studies, either when new data or more precise 
estimations of the Milan factor become available.
\section{Fit procedure and systematic uncertainties}
\label{sec:fit}
The values of $\alpha_{S}$  were determined in the optimization procedures using the 
MINUIT2~\cite{minuit,James:2004xla} program and the minimized function
\begin{equation}
\chi^2(\alpha_{S})=\sum^{\rm all\ data\ sets}_{i}\chi^2_{i}(\alpha_{S}), 
\label{chi2}
\end{equation}
where for data set $i$ we have  
$\chi^2_{i}(\alpha_{S})=(\vec{D}-\vec{P}(\alpha_{S}))V^{-1}(\vec{D}-\vec{P}(\alpha_{S}))^{T}$,
with $\vec{D}$ standing for the vector of data points, $\vec{P}(\alpha_{S})$
for the vector of calculated predictions and $V$ for the covariance
matrix of $\vec{D}$. In this analysis the covariance matrix $V$ was diagonal with  
the values of the diagonal elements calculated by adding the    
statistical and systematic uncertainties in quadrature for every measurement. 

For all the fits with MC hadronization models the central results for $\alpha_{S}(M_Z)$ 
(as well as the $D^{\langle O^n \rangle}$ coefficients in the N${}^3$LO fits) were extracted 
with the $H^{L}$ setup. The uncertainty on the fit result was estimated using the 
$\chi^2+1$ criterion as implemented in MINUIT2 ($exp.$). The systematic effects 
related to the modeling of hadronization with MCEGs were estimated as the 
difference of results obtained with $H^{L}$ and $H^{C}$ setups ($hadr.$). To 
estimate the systematic effects related to the choice of renormalization scale, 
the latter was varied  by a factor of two in both directions ($scale$). The scale 
variation at N${}^3$LO was performed with the perturbative coefficients 
$D^{\langle O^n \rangle}$ fixed to their values obtained in the nominal fit.
The uncertainty was estimated as half of the difference between the maximal and 
minimal obtained values among the three results.

When using the analytic hadronization model (with any of the three schemes discussed 
above), in addition to $\alpha_{S}(M_Z)$ and $D^{\langle O^n \rangle}$, the quantities 
$\alpha_{0}(\mu_I)$ and $\mathcal{M}$ were also treated as fit parameters to be extracted 
from data. Although the value of the Milan factor is in principle fixed by a theoretical 
calculation, including it as a constrained parameter in the fit provides a way of taking 
into account its uncertainty. For the fits with a constrained Milan factor a term 
$(\mathcal{M}-\mathcal{M}_{\rm est.})^2/(\delta\mathcal{M}_{\rm est.})^2$ is added to eq.~(\ref{chi2}).

As previously, the $\chi^2+1$ criterion was used to estimate the fit uncertainty 
($exp.$), while the systematic effects of missing higher-order terms were estimated using 
the same renormalization scale variation procedure as for MC hadronization models ($scale.$). 
In particular, when varying the scale  to estimate the related $\alpha_{S}(M_Z)$ ($\alpha_{0}(\mu_I)$) uncertainties, 
the $D^{\langle O^n \rangle}$ coefficients, 
$\alpha_{0}(\mu_I)$ ($\alpha_{S}(M_Z)$) and $\mathcal{M}$ were fixed to the values obtained in the nominal 
fits.
\section{Results and discussion}
\label{sec:results}
\begin{figure}[htbp]\centering
\includegraphics[width=\BROADFIGWIDTH]{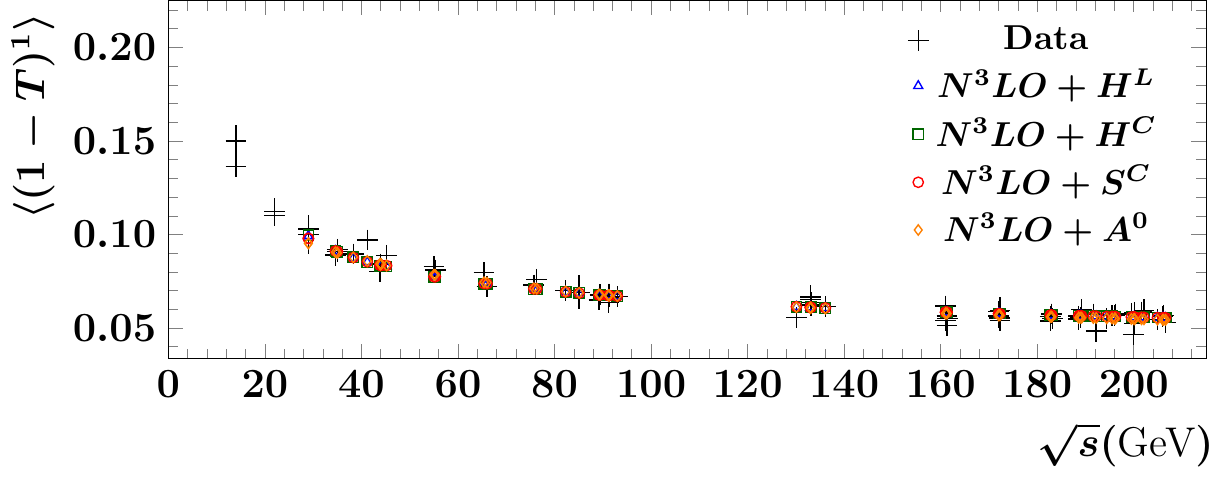}\\
\includegraphics[width=\BROADFIGWIDTH]{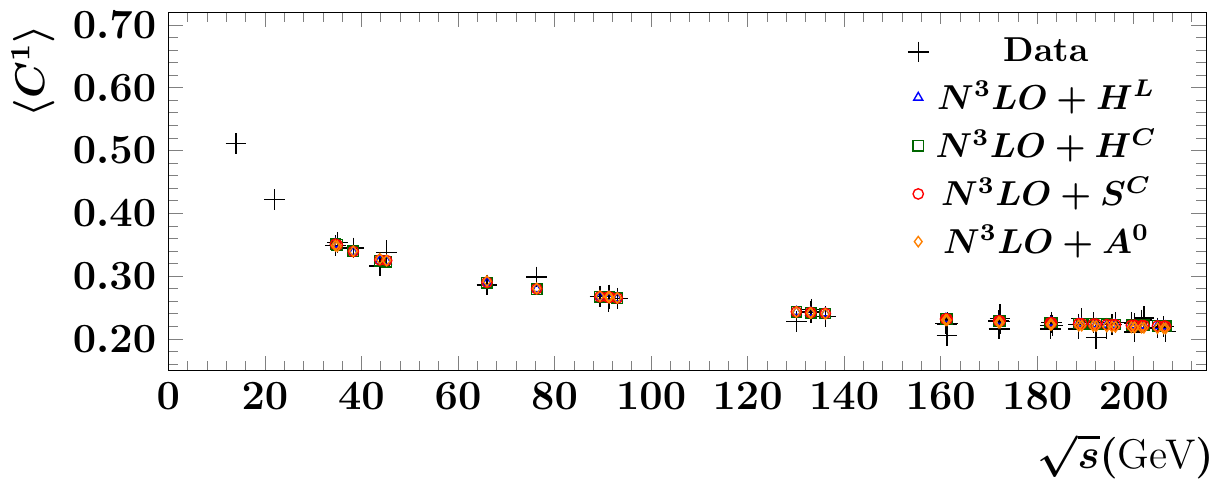}\\
\caption{Data and fits obtained with different hadronization models. All available data points from Tab.~\ref{tab:data} are shown.}
\label{fig:result}
\end{figure}

The results of the NNLO and N${}^3$LO fits are presented in Tabs.~\ref{tab:resone} 
and~\ref{tab:restwo},  while the predictions of the N$^{3}$LO fits for individual energy 
points are shown in Fig.~\ref{fig:result}.

\begin{table}[htbp]\centering
\begin{tabular}{|c|c|}\hline
Analysis             &  Results  from analysis of $\langle (1-T)^1 \rangle$ data                                      
\\\hline\hline
\epjconly{         
 NNLO,             & $\alpha_{S}(M_{Z})=\TTMOMresultNNLOpI$              \\
 MC had.           &                    $\pm\TTMOMresultNNLOpII$         \\
 $H^L$             & $\chi^2/ndof=\TTMOMresultNNLOquality$               \\\hline\hline
}
\draftonly{ 
NNLO,              & $\alpha_{S}(M_{Z})=\TTMOMresultNNLO$                \\
MC had.            & $\chi^2/ndof=\TTMOMresultNNLOquality$               \\
$H^L$              &                                                     \\\hline\hline
}
\arxivonly{ 
NNLO,              & $\alpha_{S}(M_{Z})=\TTMOMresultNNLO$                \\
MC had.            & $\chi^2/ndof=\TTMOMresultNNLOquality$               \\
$H^L$              &                                                     \\\hline\hline
}

\epjconly{
                   & $\alpha_{S}(M_{Z})=\TTbMOMresultNNLOpI$             \\
 NNLO,             &                    $\pm\TTbMOMresultNNLOpII$        \\
}
\draftonly{
 NNLO,             & $\alpha_{S}(M_{Z})=\TTbMOMresultNNLO$               \\
}
\arxivonly{
 NNLO,             & $\alpha_{S}(M_{Z})=\TTbMOMresultNNLO$               \\
}
 analytic            & $\chi^2/ndof=\TTbMOMresultNNLOquality$            \\
 had., $A^0$         & $\alpha_{0}(2\GeV)=\TTbMOMresultNNLOalphazero$    \\
                     & $\mathcal{M}=\TTbMOMresultNNLOMilan$(constrained) \\\hline\hline

\epjconly{
                     & $\alpha_{S}(M_{Z})=\TTYMOMresultNNLOpI$           \\
 NNLO,               &                   $\pm\TTYMOMresultNNLOpII$       \\
}
\draftonly{
 NNLO,               & $\alpha_{S}(M_{Z})=\TTYMOMresultNNLO$             \\
}
\arxivonly{
 NNLO,               & $\alpha_{S}(M_{Z})=\TTYMOMresultNNLO$             \\
}
 analytic            & $\chi^2/ndof=\TTYMOMresultNNLOquality$            \\
 had., $A^T$         & $\alpha_{0}(2\GeV)=\TTYMOMresultNNLOalphazero$    \\
                     & $\mathcal{M}=\TTYMOMresultNNLOMilan$(constrained) \\\hline\hline

\epjconly{
                     & $\alpha_{S}(M_{Z})=\TTYMOMresultNNLOpI$           \\
 NNLO,               &                   $\pm\TTBMOMresultNNLOpII$       \\
}
\draftonly{
 NNLO,               & $\alpha_{S}(M_{Z})=\TTBMOMresultNNLO$             \\
}
\arxivonly{
 NNLO,               & $\alpha_{S}(M_{Z})=\TTBMOMresultNNLO$             \\
}
 analytic            & $\chi^2/ndof=\TTBMOMresultNNLOquality$            \\
 had., $A^{\mathrm{cusp}}$         & $\alpha_{0}(2\GeV)=\TTBMOMresultNNLOalphazero$    \\
                     & $\mathcal{M}=\TTBMOMresultNNLOMilan$(constrained) \\\hline\hline

\epjconly{
 N${}^3$LO,       & $\alpha_{S}(M_{Z})=\TTMOMresultNNNLOpI$              \\
 MC had.          &                    $\pm\TTMOMresultNNNLOpII$         \\
 $H^L$            & $\chi^2/ndof=\TTMOMresultNNNLOquality$               \\
                  & $D^{\langle (1-T)^1 \rangle}=\TTDoneminusTpowone$    \\\hline\hline
}
\draftonly{
 N${}^3$LO,          & $\alpha_{S}(M_{Z})=\TTMOMresultNNNLO$             \\
 MC had.             & $\chi^2/ndof=\TTMOMresultNNNLOquality$            \\
 $H^L$               & $D^{\langle (1-T)^1 \rangle}=\TTDoneminusTpowone$ \\\hline\hline
}
\arxivonly{
 N${}^3$LO,          & $\alpha_{S}(M_{Z})=\TTMOMresultNNNLO$             \\
 MC had.             & $\chi^2/ndof=\TTMOMresultNNNLOquality$            \\
 $H^L$               & $D^{\langle (1-T)^1 \rangle}=\TTDoneminusTpowone$ \\\hline\hline
}

\epjconly{                     
                     & $\alpha_{S}(M_{Z})=\TTbMOMresultNNNLOpI$          \\
 N${}^3$LO,          &                 $\pm\TTbMOMresultNNNLOpII$        \\
}
\draftonly{                     
 N${}^3$LO,          & $\alpha_{S}(M_{Z})=\TTbMOMresultNNNLO$            \\
}
\arxivonly{                     
 N${}^3$LO,          & $\alpha_{S}(M_{Z})=\TTbMOMresultNNNLO$            \\
}
 analytic            & $\chi^2/ndof=\TTbMOMresultNNNLOquality$           \\
 had., $A^0$         & $D^{\langle (1-T)^1 \rangle}=\TTbDoneminusTpowone$\\
                     & $\alpha_{0}(2\GeV)=\TTbMOMresultNNNLOalphazero$   \\
                     & $\mathcal{M}=\TTbMOMresultNNNLOMilan$(constrained)\\\hline\hline
\epjconly{ 
                     & $\alpha_{S}(M_{Z})=\TTYMOMresultNNNLOpI$          \\
 N${}^3$LO,          &                 $\pm\TTYMOMresultNNNLOpII$        \\
}
\draftonly{ 
 N${}^3$LO,          & $\alpha_{S}(M_{Z})=\TTYMOMresultNNNLO$            \\
}
\arxivonly{ 
 N${}^3$LO,          & $\alpha_{S}(M_{Z})=\TTYMOMresultNNNLO$            \\
}

 analytic            & $\chi^2/ndof=\TTYMOMresultNNNLOquality$           \\
 had., $A^T$         & $D^{\langle (1-T)^1 \rangle}=\TTYDoneminusTpowone$ \\
                     & $\alpha_{0}(2\GeV)=\TTYMOMresultNNNLOalphazero$   \\
                     & $\mathcal{M}=\TTYMOMresultNNNLOMilan$(constrained) \\\hline\hline
\epjconly{ 
                     & $\alpha_{S}(M_{Z})=\TTBMOMresultNNNLOpI$          \\
 N${}^3$LO,          &                $\pm\TTBMOMresultNNNLOpII$         \\
}
\draftonly{ 
 N${}^3$LO,          & $\alpha_{S}(M_{Z})=\TTBMOMresultNNNLO$            \\
}
\arxivonly{ 
 N${}^3$LO,          & $\alpha_{S}(M_{Z})=\TTBMOMresultNNNLO$            \\
}
 analytic            & $\chi^2/ndof=\TTBMOMresultNNNLOquality$           \\
 had., $A^{\mathrm{cusp}}$         & $D^{\langle (1-T)^1 \rangle}=\TTBDoneminusTpowone$ \\
                     & $\alpha_{0}(2\GeV)=\TTBMOMresultNNNLOalphazero$   \\
                     & $\mathcal{M}=\TTBMOMresultNNNLOMilan$(constrained) \\\hline

\end{tabular}
\caption{Results of the extraction analyses using the $\langle (1-T)^1 \rangle$ observable.}    
\label{tab:resone}
\end{table}

\begin{table}[htbp]\centering
\begin{tabular}{|c|c|}\hline
Analysis             &  Results   from analysis of $\langle C^1\rangle$ data \\\hline\hline

\epjconly{ 
 NNLO,               & $\alpha_{S}(M_{Z})=\CCMOMresultNNLOpI$            \\
 MC had.             &                $\pm\CCMOMresultNNLOpII$           \\
 $H^L$               & $\chi^2/ndof=\CCMOMresultNNLOquality$             \\\hline\hline
} 
\draftonly{ 
 NNLO,               & $\alpha_{S}(M_{Z})=\CCMOMresultNNLO$              \\
 MC had.             & $\chi^2/ndof=\CCMOMresultNNLOquality$             \\
 $H^L$               &                                                   \\\hline\hline
} 
\arxivonly{ 
 NNLO,               & $\alpha_{S}(M_{Z})=\CCMOMresultNNLO$              \\
 MC had.             & $\chi^2/ndof=\CCMOMresultNNLOquality$             \\
 $H^L$               &                                                   \\\hline\hline
}

\epjconly{ 
                     & $\alpha_{S}(M_{Z})=\CCbMOMresultNNLOpI$           \\
 NNLO,               &                 $\pm\CCbMOMresultNNLOpII$         \\
}                     
\draftonly{ 
 NNLO,               & $\alpha_{S}(M_{Z})=\CCbMOMresultNNLO$             \\
}                     
\arxivonly{ 
 NNLO,               & $\alpha_{S}(M_{Z})=\CCbMOMresultNNLO$             \\
}
 analytic            & $\chi^2/ndof=\CCbMOMresultNNLOquality$            \\
 had., $A^0$         & $\alpha_{0}(2\GeV)=\CCbMOMresultNNLOalphazero$    \\
                     & $\mathcal{M}=\CCbMOMresultNNLOMilan$(constrained) \\\hline\hline

\epjconly{ 
                     & $\alpha_{S}(M_{Z})=\CCYMOMresultNNLOpI$           \\
 NNLO,               &                  $\pm\CCYMOMresultNNLOpII$        \\
}
\draftonly{ 
 NNLO,               & $\alpha_{S}(M_{Z})=\CCYMOMresultNNLO$             \\
}
\arxivonly{ 
 NNLO,               & $\alpha_{S}(M_{Z})=\CCYMOMresultNNLO$             \\
}
 analytic            & $\chi^2/ndof=\CCYMOMresultNNLOquality$            \\
 had. $A^T$          & $\alpha_{0}(2\GeV)=\CCYMOMresultNNLOalphazero$    \\
                     & $\mathcal{M}=\CCYMOMresultNNLOMilan$(constrained) \\\hline\hline

\epjconly{
                     & $\alpha_{S}(M_{Z})=\CCBMOMresultNNLOpI$           \\
 NNLO,               &                $\pm\CCBMOMresultNNLOpII$          \\
}
\draftonly{
 NNLO,               & $\alpha_{S}(M_{Z})=\CCBMOMresultNNLO$             \\
}
\arxivonly{
 NNLO,               & $\alpha_{S}(M_{Z})=\CCBMOMresultNNLO$             \\
}
 analytic            & $\chi^2/ndof=\CCBMOMresultNNLOquality$            \\
 had., $A^{\mathrm{cusp}}$         & $\alpha_{0}(2\GeV)=\CCBMOMresultNNLOalphazero$ \\
                     & $\mathcal{M}=\CCBMOMresultNNLOMilan$(constrained) \\\hline\hline

\epjconly{
 N${}^3$LO,         & $\alpha_{S}(M_{Z})=\CCMOMresultNNNLOpI$            \\
 MC had.            &                $\pm\CCMOMresultNNNLOpII$           \\
 $H^L$              & $\chi^2/ndof=\CCMOMresultNNNLOquality$             \\
                    & $D^{\langle C^1 \rangle}=\CCDCpowone$              \\\hline\hline
}                     
\draftonly{
 N${}^3$LO,          & $\alpha_{S}(M_{Z})=\CCMOMresultNNNLO$             \\
 MC had.             & $\chi^2/ndof=\CCMOMresultNNNLOquality$            \\
 $H^L$               & $D^{\langle C^1 \rangle}=\CCDCpowone$             \\\hline\hline
}                     
\arxivonly{
 N${}^3$LO,          & $\alpha_{S}(M_{Z})=\CCMOMresultNNNLO$             \\
 MC had.             & $\chi^2/ndof=\CCMOMresultNNNLOquality$            \\
 $H^L$               & $D^{\langle C^1 \rangle}=\CCDCpowone$             \\\hline\hline
}

\epjconly{ 
                      & $\alpha_{S}(M_{Z})=\CCbMOMresultNNNLOpI$         \\
  N${}^3$LO,          &                $\pm\CCbMOMresultNNNLOpII$        \\
}
\arxivonly{  
 N${}^3$LO,          & $\alpha_{S}(M_{Z})=\CCbMOMresultNNNLO$            \\
}
\draftonly{
 N${}^3$LO,          & $\alpha_{S}(M_{Z})=\CCbMOMresultNNNLO$            \\
}
 analytic            & $\chi^2/ndof=\CCbMOMresultNNNLOquality$           \\
 had. $A^0$          & $D^{\langle C^1 \rangle}=\CCbDCpowone$            \\
                     & $\alpha_{0}(2\GeV)=\CCbMOMresultNNNLOalphazero$   \\
                     & $\mathcal{M}=\CCbMOMresultNNNLOMilan$(constrained)\\\hline\hline

\epjconly{ 
                     & $\alpha_{S}(M_{Z})=\CCYMOMresultNNNLOpI$          \\
 N${}^3$LO,          &                 $\pm\CCYMOMresultNNNLOpII$        \\
}
\arxivonly{ 
 N${}^3$LO,          & $\alpha_{S}(M_{Z})=\CCYMOMresultNNNLO$            \\
} 
\draftonly{ 
 N${}^3$LO,          & $\alpha_{S}(M_{Z})=\CCYMOMresultNNNLO$            \\
}
 analytic            & $\chi^2/ndof=\CCYMOMresultNNNLOquality$           \\
 had. $A^T$          & $D^{\langle C^1 \rangle}=\CCYDCpowone$            \\
                     & $\alpha_{0}(2\GeV)=\CCYMOMresultNNNLOalphazero$   \\
                     & $\mathcal{M}=\CCYMOMresultNNNLOMilan$(constrained)\\\hline\hline

\epjconly{ 
                     & $\alpha_{S}(M_{Z})=\CCBMOMresultNNNLOpI$          \\
 N${}^3$LO,          &                $\pm\CCBMOMresultNNNLOpII$         \\
}
\draftonly{ 
 N${}^3$LO,          & $\alpha_{S}(M_{Z})=\CCBMOMresultNNNLO$            \\
}
\arxivonly{ 
 N${}^3$LO,          & $\alpha_{S}(M_{Z})=\CCBMOMresultNNNLO$            \\
}
 analytic            & $\chi^2/ndof=\CCBMOMresultNNNLOquality$           \\
 had. $A^{\mathrm{cusp}}$          & $D^{\langle C^1 \rangle}=\CCBDCpowone$ \\
                     & $\alpha_{0}(2\GeV)=\CCBMOMresultNNNLOalphazero$   \\
                     & $\mathcal{M}=\CCBMOMresultNNNLOMilan$(constrained) \\\hline

\end{tabular}
\caption{Results of the extraction analyses using the $\langle C^1 \rangle$ observable.}    
\label{tab:restwo}
\end{table}
\draftonly{\FloatBarrier}
\arxivonly{\FloatBarrier}
The presented NNLO results for $\alpha_{S}(M_Z)$ obtained with both MC and analytic 
hadronization models are in good agreement between the fits to $\langle (1-T)^1 \rangle$ 
and $\langle C^1 \rangle$, which can be viewed as a check of the internal consistency of 
the $\alpha_{S}(M_{Z})$ extraction method at NNLO. However, similarly to previous 
studies~\cite{Gehrmann:2009eh} which used less data\footnote{The analysis of 
Ref.~\cite{Gehrmann:2009eh} employed several other event shape variables besides 
thrust and the $C$-parameter (as well as higher moments of event shapes), but the 
present study uses a more extensive data set for the observables considered here.}, 
a large discrepancy between the results obtained with the MC hadronization model and 
the analytic hadronization models are seen.\footnote{In previous studies~\cite{Gehrmann:2009eh} 
the results obtained with the MC hadronization model were systematically higher 
than those obtained with the analytic hadronization model, while in the presented 
study an opposite relation is seen. This difference can be attributed to differences 
in the used data sets and MC setups.}

Turning to $\alpha_0(2\GeV)$ still at NNLO, we recall that this parameter is 
scheme-dependent, so the fitted values in the three schemes should not be directly 
compared to each other. Nevertheless, we see that the choice of scheme has only a 
small numerical impact on the extracted values of $\alpha_0(2\GeV)$. The values of 
the Milan parameter $\mathcal{M}$, constrained in fits, are seen to be unaffected 
by the choice of scheme and agree with the theoretical prediction within the somewhat 
large fit uncertainty. Furthermore, the extracted values of both $\alpha_0(2\GeV)$ and 
$\mathcal{M}$ obtained form the $\langle (1-T)^1 \rangle$ and $\langle C^1 \rangle$ 
observables agree well with each other.

\begin{figure}[htbp]\centering
\includegraphics[width=\NARROWFIGWIDTH]{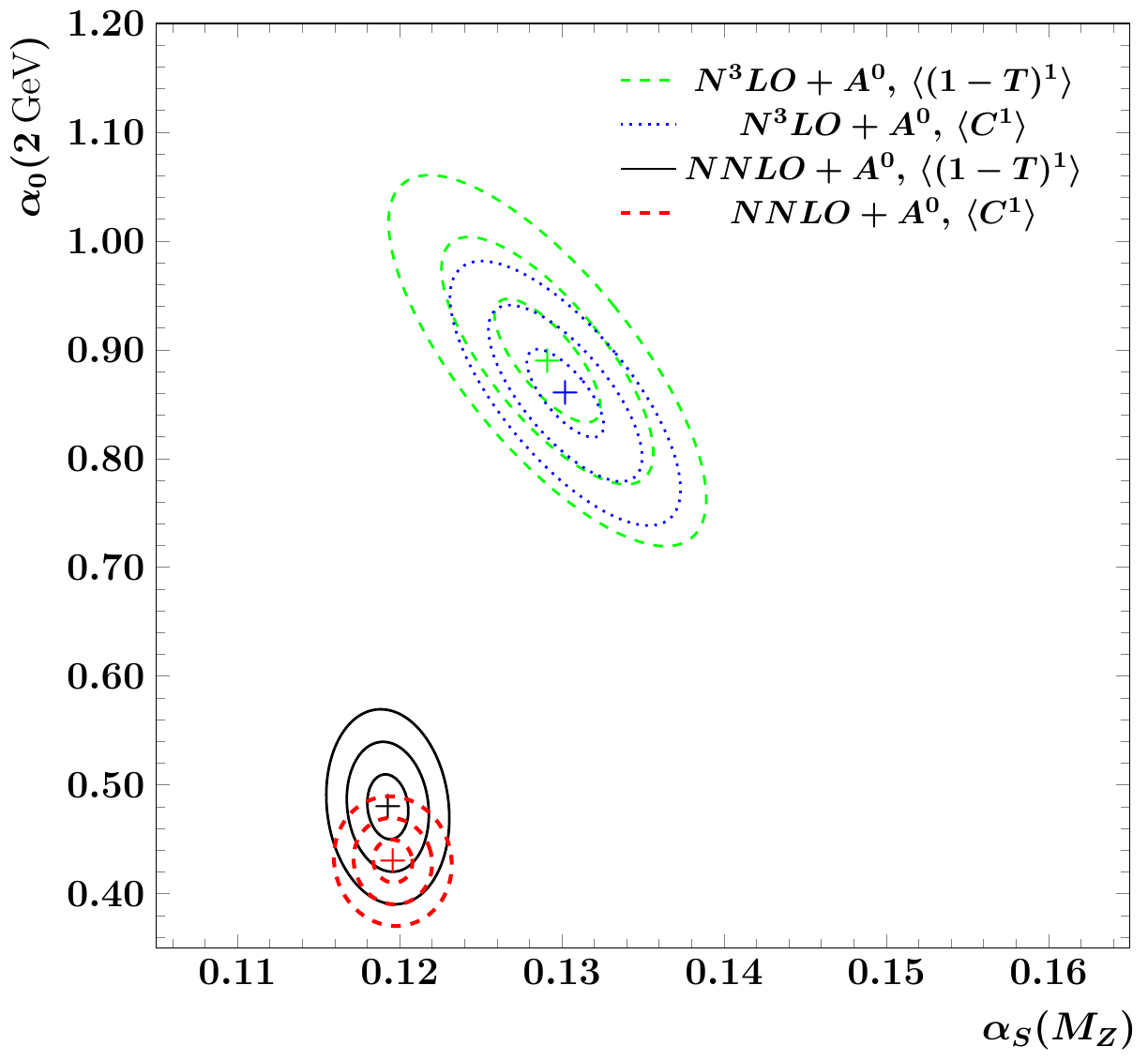}\\
\caption{The values of $\alpha_{S}(M_{Z})$ and $\alpha_{0}(2\GeV)$ obtained from 
the NNLO and N$^{3}$LO fits with analytic hadronization model in the $\mathcal{A}^0$ 
scheme. The contours correspond to $1$-, $2$- and $3$ standard deviations obtained 
in the fit. Systematic uncertainties are not included.}
\label{fig:corrplot}
\end{figure}

Turning to the N${}^{3}$LO results, we see that the overall picture is quite similar 
to the one at NNLO: the fits for $\alpha_{S}(M_Z)$ are in good agreement between the 
two observables for both MC and analytic hadronization models. The extracted values 
of $\alpha_0(2\GeV)$ and $\mathcal{M}$ are also consistent between the determinations 
based on $\langle (1-T)^1 \rangle$ and $\langle C^1 \rangle$. However, for all of these quantities we find rather 
large uncertainties, primarily related to the insufficient amount and quality of data 
and the extraction method itself. Nevertheless, these uncertainties are not very 
much larger than those from some classical $\alpha_{S}(M_{Z})$ extraction 
analyses in the past~\cite{Schieck:2012mp}. 
Moreover, the obtained values of both $D^{\langle (1-T)^1 \rangle}$ and 
$D^{\langle C^1 \rangle}$ are in reasonable agreement between fits using MC and analytic 
hadronization models. This demonstrates the viability of the extraction of 
the higher-order coefficients $D^{\langle O^n \rangle}$, once a large amount 
of precise and consistent data will be available, e.g.\  from 
CEPC~\cite{CEPCStudyGroup:2018ghi} or FCC-$e e$~\cite{Abada:2019zxq}.
 For this extraction the precise high-energy data would be especially valuable.
Finally, in Fig.~\ref{fig:corrplot} we present the extracted values of  
$\alpha_{S}(M_{Z})$ and $\alpha_{0}(2\GeV)$ at NNLO and N${}^3$LO accuracy in the 
$\mathcal{A}_0$-scheme. The results at each perturbative order are quite consistent 
across the two observables and the fits for $\langle (1-T)^1 \rangle$ and 
$\langle C^1 \rangle$ have rather similar precision. However, the fits at  N${}^3$LO 
clearly prefer larger values for both $\alpha_{S}(M_{Z})$ and $\alpha_{0}(2\GeV)$.
At the same time, the discrepancy between results obtained with the MC 
hadronization model and the analytic hadronization model remains in place 
at N${}^{3}$LO accuracy. This suggests that the discrepancy pattern has a 
fundamental origin and would hold even in future analyses, regardless of 
the availability of the exact N${}^{3}$LO predictions. Consequently, the 
improvement of the hadronization modeling and a better understanding of 
hadronization itself is more important for increasing the precision of 
$\alpha_{S}(M_{Z})$ extractions than the calculation of perturbative 
corrections beyond NNLO. In order to achieve this better understanding 
and improved modeling of hadronization, in the future it would be important 
to perform dedicated studies using observables strongly affected by 
hadronization, e.g.\ measurements of the hadronic final state in 
future $e^{+}e^{-}$ experiments at $\sqrt{s}\approx 20-50\GeV$ performed 
with radiative events or in dedicated collider runs.

\section{Conclusions}
\label{sec:conclusions}
 The aim of the present analysis was to assess the factors that will determine the  
precision of QCD analyses of $e^{+}e^{-}$ data after
 the recent and foreseen rapid developments of QCD calculation techniques and the appearance
 of even more precise theoretical predictions. To do this, we
have performed an extraction of the strong coupling $\alpha_{S}(M_{Z})$ from 
 the averages of event shapes event shape moments
 $\langle (1-T)^1 \rangle$ and $\langle C^1 \rangle$ and found that 
the results obtained using NNLO predictions and analytic hadronization 
corrections based on the dispersive model are consistent with the last PDG 
average $\alpha(M_Z)_{\rm PDG 2020}  = 0.1179 \pm 0.0010$.

Furthermore, we considered a method for extracting $\alpha_{S}(M_{Z})$ at 
N${}^3$LO precision in perturbative QCD, employing exact NNLO predictions 
and estimations of the N${}^3$LO corrections from the data. 
The method produced results which are compatible with the current world average 
within the somewhat large uncertainties,
e.g.\  $$\alpha_{S}(M_{Z})^{N^{3}LO+A^0}=\TTbMOMresultNNNLO$$ from the $\langle (1-T)^1 \rangle$ data.
The obtained precision can be increased with more high-quality data from future experiments. 
For the extraction, Monte Carlo and analytic hadronization 
models were used, the latter being extended to N${}^3$LO for the first time. 
The comparison of the results for these models suggests that extractions of 
$\alpha_{S}(M_{Z})$ in future analyses will be strongly affected by the modeling 
of hadronization effects even  when the exact higher-order corrections will be included.
However, the improvements in the modeling of high-energy physics phenomena by MCEG in 
the recent decades were closely tied to the experimental measurements performed at the 
LEP, HERA and LHC colliders and therefore had limited impact on the description of 
phenomena at lower energies. As a consequence, the advances in modeling of particle 
collisions at lower energies  and understanding of hadronization can be expected only 
with the availability of new measurements in the corresponding energy ranges.

\section*{Acknowledgements}
\label{sec:acknowledgements}
We are grateful to Simon Pl\"{a}tzer for fruitful discussions about the 
calculation of NLO predictions with {\tt Herwig7.2.0} and to Carlo 
Oleari for providing  us the \prog{Zbb4} code. 
We are grateful to Daniel Britzker, Stefan Kluth and Pier Monni for the discussions 
on the analysis.

A.K. acknowledges financial support from the Premium Postdoctoral 
Fellowship program of the Hungarian Academy of Sciences. This work was 
supported by grant K 125105 of the National Research, Development and
Innovation Fund in Hungary.
\FloatBarrier

\appendix
\newpage 
\allowdisplaybreaks
\section{Perturbative coefficients $A_{\mathrm{tot}}$, $B_{\mathrm{tot}}$ and $C_{\mathrm{tot}}$}
\label{app:ABCtot}
In this appendix, we recall the total cross section, $\sigma_{\mathrm{tot}}$, of 
electron-positron annihilation into hadrons. In massless QCD with $N_F$ number of 
light flavors we have~\cite{Baikov:2012zn},
\begin{align}
\sigma_{\mathrm{tot}} = \sigma_{0}
	\left[1 
	+ \left(\frac{\alpha_{S}}{2\pi}\right)^1 A_{\mathrm{tot}}
	\right.&+ \left(\frac{\alpha_{S}}{2\pi}\right)^2 B_{\mathrm{tot}}
	+ \epjconly{ \nonumber\\ & +}\left. \left(\frac{\alpha_{S}}{2\pi}\right)^3 C_{\mathrm{tot}}
	+ \mathcal{O}(\alpha_{S}^4)\right]\,,
\end{align}
with
\begin{align}
A_{\mathrm{tot}} &= \frac{3}{2}C_F\,,
\nonumber
\\
B_{\mathrm{tot}} &= C_F\left[
    \left(\frac{123}{8} - 11\zeta_3\right)C_A 
    - \frac{3}{8} C_F
    -\left(\frac{11}{4} - 2\zeta_3\right) N_F\right]\,,
\nonumber    
\\
C_{\mathrm{tot}} &= C_F \bigg[
    \left(\frac{90445}{432} - \frac{2737}{18} \zeta_3 - \frac{55}{3} \zeta_5\right) C_A^2 \epjconly{\notag\\&}
    - \left(\frac{127}{8} + \frac{143}{2} \zeta_3 - 110 \zeta_5\right) C_A C_F 
    -\frac{69}{16} C_F^2  
\notag\\&
    - \left(\frac{1940}{27} - \frac{448}{9} \zeta_3 - \frac{10}{3} \zeta_5\right) C_A N_F \epjconly{\notag\\&}
    - \left(\frac{29}{16} - 19 \zeta_3 + 20 \zeta_5\right) C_F N_F 
\notag\\&
    + \left(\frac{151}{27} - \frac{38}{9} \zeta_3\right) N_F^2 
    - \frac{\pi^2}{8} \left(\frac{11}{3} C_A - \frac{2}{3} N_F\right)^2\bigg] \epjconly{\notag\\&}
    + \frac{(\sum Q_f)^2}{3\sum Q_f^2} \frac{d^{abc}d^{abc}}{16} 
    \left(\frac{22}{3} - 16 \zeta_3\right)\,.
\nonumber
\end{align}
We recall that we use $T_R=1/2$ and so $C_A = N_c = 3$, $C_F = (N_c^2-1)/(2N_c) = 4/3$ 
and $d^{abc}d^{abc} = 40/3$. Furthermore, $Q_f$ denotes the electric charge 
of quarks and $N_F$ is the number of light quark flavors.

\section{Analytic calculations of  the perturbative coefficients for the event shape moments}
\label{app:AB}
The LO analytic results for 
$\langle (1-T)^n \rangle$, i.e.\ $A_0^{\langle (1-T)^n \rangle}$  use the calculations from Ref.~\cite{DeRujula:1978vmq} and read

\begin{align*}
A_0^{\langle (1-T)^1 \rangle}&= C_F \left( -\frac{3}{4}\ln(3)   - \frac{1}{18}   +\epjconly{\right.\notag\\&\left.+} \frac{\pi^2}{3} + 4\mathrm{Li}_2(3/2)+2\ln(2)^2\right)=2.1034701\ldots,
\nonumber \\
A_0^{\langle (1-T)^2 \rangle}&= C_F \left( -\frac{9}{4}\ln(3)   + \frac{17}{108} +\epjconly{\right.\notag\\&\left.+} \frac{\pi^2}{3} + 4\mathrm{Li}_2(3/2)+2\ln(2)^2\right)=0.1901961\ldots,
\nonumber \\
A_0^{\langle (1-T)^3 \rangle}&= C_F \left( -\frac{83}{32}\ln(3) + \frac{56}{135} +\epjconly{\right.\notag\\&\left.+} \frac{\pi^2}{3} + 4\mathrm{Li}_2(3/2)+2\ln(2)^2\right)=0.0298753\ldots,
\nonumber \\
A_0^{\langle (1-T)^4 \rangle}&= C_F \left( -\frac{649}{240}\ln(3) + \frac{1259}{2430} +\epjconly{\right.\notag\\&\left.+} \frac{\pi^2}{3} + 4\mathrm{Li}_2(3/2)+2\ln(2)^2\right)=0.0058581\ldots,
\nonumber \\
A_0^{\langle (1-T)^5 \rangle}&= C_F \left( -\frac{527}{192}\ln(3) + \frac{45667}{81648} +\epjconly{\right.\notag\\&\left.+} \frac{\pi^2}{3} + 4\mathrm{Li}_2(3/2)+2\ln(2)^2\right)=0.0012947\ldots,
\nonumber \\
\end{align*}

The analytic result for $A_0^{\langle C^1 \rangle}$ has been known for a long 
time~\cite{Ellis:1980wv}. This result, same as the results for the 
higher moments can be also obtained with a direct integration, e.g.\ using the calculations from Ref.~\cite{Preisser:2014}. These read:
\begin{align*}
A_0^{\langle C^1 \rangle} &=C_F \left(-33 + 4 \pi^2\right)=8.6378901\ldots\,,
\nonumber \\
A_0^{\langle C^2 \rangle} &=C_F \left( 594 - 60 \pi^2\right)=2.4316479\ldots\,,
\nonumber \\
A_0^{\langle C^3 \rangle} &=C_F \left(-6750 + 684 \pi^2\right)=1.0792137\ldots\,,
\nonumber \\
A_0^{\langle C^4 \rangle} &=C_F \left( 65088 - \frac{26379}{4} \pi^2\right)=0.5685012\ldots\,,
\nonumber \\
A_0^{\langle C^5 \rangle} &=C_F \left(-570024 + \frac{1848177}{32} \pi^2\right)=0.327216\ldots\,,
\nonumber \\
\end{align*}
 The NLO coefficient $B_0^{\langle C^1 \rangle}$ was calculated  for the first time
using the analytic expression for the energy-energy correlations (EEC) from 
Ref.~\cite{Dixon:2018qgp} and the identity on the event level 
$\langle C^1 \rangle =\frac{3}{2}\int^{1}_{-1} EEC(\theta)  \sin^2{\theta} d(cos\theta)$ 
and we find:
\begin{align*}
B_0^{\langle C^1 \rangle} &= C_F N_F T_R \left(\frac{18759}{140} - 7\pi^2 - \frac{2728 \zeta_3}{35}\right) 
\epjconly{\\&} + C_F^2 \left(-\frac{8947}{224} + \frac{101 \pi^2}{24} + \frac{2 \pi^4}{15} - \frac{201 \zeta_3}{7}\right) 
\draftonly{\notag\\&}
\arxivonly{\notag\\&}
\epjconly{\notag\\&}
 + C_A C_F \left(-\frac{209821}{840} + \frac{247 \pi^2}{18} - \frac{8 \pi^4}{15} + \frac{ 7057 \zeta_3}{35}\right)
 \epjconly{\notag\\&}=172.85901\ldots\,.
\nonumber
\end{align*}

The results that account for non-zero quark masses are not known analytically even at LO, however
the coefficient
$A^{\langle C^{1} \rangle}_{m_b\ne 0}$ could be derived in a closed from using the results for 
$EEC_{m_b\ne 0}$ from Refs.~\cite{Cho:1984rq,Csikor:1983dt} or the
results for the $\frac{d C}{d\sigma}\big|_{m_b\ne 0}$ from Ref.~\cite{Preisser:2014}.

\section{The $K$, $L$ and $M$ coefficients in different schemes}
\label{app:KLM}
The $K$, $L$ and $M$ coefficients in the cusp-scheme are simply given 
by the one-, two- and three-loop cusp anomalous dimensions (for quarks) 
and read~\cite{Curci:1980uw,Furmanski:1980cm,Moch:2004pa,Vogt:2004mw,Moch:2017uml,Moch:2018wjh}:
\begin{align*}
K_{\mathrm{cusp}}&=C_A\bigg(\frac{67}{18}-\frac{\pi^{2}}{6}\bigg)-\frac{5}{9}N_F,\\
L_{\mathrm{cusp}}&=C^2_A\bigg(\frac{245}{24}-\frac{67\pi^2}{54}+\frac{11\zeta_3}{6}+\frac{11\pi^4}{180}\bigg) 
\epjconly{\nonumber \\&} + C_F N_F\bigg(-\frac{55}{24}+2\zeta_3\bigg)
\nonumber \\ &
+C_A N_F\bigg(-\frac{209}{108}+\frac{5\pi^2}{27}-\frac{7\zeta_3}{3}\bigg)-\frac{1}{27}N^2_F,\\
M_{\mathrm{cusp}}&=\frac{3}{128}(20702 - 5171.9 N_F + 195.5772 N_F^2 \epjconly{\nonumber \\&} +  3.272344 N_F^3).
\end{align*}

The $K$, $L$ and $M$ coefficients in the $\mathcal{A}_{0}$-scheme read~\cite{Banfi:2018mcq,Catani:2019rvy}:
\begin{align*}
K_{0}&=K_{\mathrm{cusp}},\\
L_{0}&=L_{\mathrm{cusp}}+C_A^2 \left(     \frac{77\zeta_3}{6} -\frac{1111}{81} \right)
 \epjconly{\nonumber \\&}+C_A N_F \left(  -\frac{11\pi^2}{27}  + \frac{356}{81} -\frac{7\zeta_3}{3} \right) 
+ N_F^2 \left( \frac{\pi^2}{27} -\frac{28}{81}\right),\\
M_{0}&=M_{\mathrm{cusp}}
+C_A^3 \left(\frac{121\pi^2 \zeta_3}{26}  - \frac{21755\zeta_3}{108}  + 66 \zeta_5 + \frac{847\pi^4}{2160}   
\right.\epjconly{\nonumber \\&}\left. - \frac{41525\pi^2}{1944}  + \frac{3761815}{23328}\right) 
\draftonly{\nonumber \\ &}
\arxivonly{\nonumber \\ &}
+C_A^2 N_F \left(-\frac{11\pi^2 \zeta_3}{18}  + \frac{6407\zeta_3}{108}  
\right.\epjconly{\nonumber \\&}\left. - 12\zeta_5  - \frac{11 \pi^4}{54} + \frac{9605\pi^2}{972}  - \frac{15593}{243}\right)
\nonumber \\ &
+C_A C_F N_F \left(\frac{136 \zeta_3}{9} + \frac{11\pi^4}{180}  + \frac{55\pi^2}{72}  - \frac{7351}{288}\right)  
\nonumber \\ &
+C_A N_F^2 \left(-\frac{179\zeta_3}{54}  + \frac{13\pi^4}{540}  - \frac{695\pi^2}{486}  + \frac{13819}{1944}\right) 
\nonumber \\ &
+C_F N_F^2 \left(-\frac{19\zeta_3}{9}  - \frac{\pi^4}{90} - \frac{5\pi^2 }{36} + \frac{215}{48}\right)  
\epjconly{\nonumber \\&} + N_F^3 \left(-\frac{2\zeta_3}{27} + \frac{5\pi^2}{81}  - \frac{116}{729}\right),
\end{align*}

The $K$, $L$ and $M$ coefficients in the $\mathcal{A}_{T}$-scheme read~\cite{Catani:2019rvy}:
\begin{align*}
K_{T}&=K_{\mathrm{cusp}},\\
L_{T}&=L_{\mathrm{cusp}}+C_A^2 \left( \frac{77\zeta_3}{6} - \frac{111}{81} \right) - C_A N_F  \left( \frac{7\zeta_3}{3}  -\frac{356}{81} \right)  
\epjconly{\nonumber \\&} -\frac{28}{81}N_F^2.\\
\end{align*}
We remind the reader that the complete expression for $M$ is currently not known in the 
$\mathcal{A}_{T}$-scheme, hence as an approximation, we set $M_T = M_0$ in this analysis.
However, we hope that in the future 
it will be possible to extend the calculations in the 
Ref.~\cite{Catani:2019rvy} to higher orders and calculate the $M_T$ explicitly.